\documentstyle[12pt,pra,aps,epsf,eqsecnum,feynmf
]{revtex}
\def\comment#1{}
\def\cm#1{}

\newcommand{\mean}[1]{\langle\,#1\,\rangle_\Omega^{x_b,x_a}}
\newcommand{\cum}[1]{\langle\,#1\,\rangle_{\Omega,c}^{x_b,x_a}}
\newcommand{\tx}{\tilde{x}}
\newcommand{\tp}{\tilde{p}}
\newcommand{\xvert}[1]{\fmfv{decor.shape=pentagram, decor.filled=1, decor.size=3thick}{#1}}
\newcommand{\pvert}[1]{\fmfv{decor.shape=pentagram, decor.filled=0, decor.size=3thick}{#1}}
\def\setval{\fmfset{wiggly_len}{2mm}\fmfset{arrow_len}{2mm}\fmfset{dash_len}{1.5mm}\fmfpen{0.125mm}\fmfset{dot_size}{1thick}}

\begin{document}
\title{
Correlation Functions of Harmonic Fluctuating Paths With Fixed End Points and 
Time-Dependent Frequency}
\author{Hagen Kleinert, Axel Pelster, and Michael Bachmann\\
        Institut f\"ur Theoretische Physik, Freie Universit\"at Berlin,\\
        Arnimallee 14, D--14195 Berlin, Germany\\ $\mbox{}$\\}
\date{\today}
\maketitle
\begin{abstract}

We introduce a general class of
generating functionals for the calculation of quantum-mechanical
expectation values of arbitrary functionals of fluctuating paths
with fixed end points in configuration or momentum space.
The generating functionals are calculated
explicitly for harmonic oscillators with time-dependent frequency,
and used to derive a smearing formulas for  correlation functions
of polynomial and nonpolynomials functions
of  time-dependent positions and momenta. These formulas
summarize the effect of thermal and quantum fluctuations,
and serve to derive generalized Wick
rules and Feynman diagrams
 for perturbation expansions of nonpolynomial interactions.
\end{abstract}
\section{Introduction}
A useful technique for describing compactly the properties of
a quantum mechanical system is to define a suitable generating functional
of some external source or current $j ( t )$. The desired properties are obtained from functional derivatives with respect to $j(t)$. For example, the correlation functions and 
the quantum mechanical density matrix 
in one space dimension $x$ is determined by a generating functional which is a path integral in configuration space over all paths $x(t)$ with fixed end points $x ( t_a ) = x_a, x ( t_b ) = x_b$ \cite[Chap.~2]{Kleinert}:
\begin{equation}
(x_b\, t_b\,|\, x_a\, t_a) [j(t)] =
\int\limits^{x_b,t_b}_{x_a,t_a} 
{\cal D} x(t)
\exp \left\{ \frac{i}{\hbar}  {\cal A} [x(t) ; j(t)]\right\} \, ,
\label{01}
\end{equation}
where the exponent contains the classical action ${\cal A} [x(t)]$
plus a source term
linear in  $x ( t )$:
\begin{equation}
{\cal A} [x(t); j(t)] = {\cal A} [x(t)] +
\int\limits^{t_b}_{t_b} dt 
\, x(t) j ( t ) \, .
\label{02}
\end{equation}
In this note we set up a useful alternative expression for the generating functional (\ref{01}) and a related one in momentum space. This alternative expression is obtained 
by extending the current $j ( t )$ by singular sources proportional to
$\dot{\delta} ( t_b 
- t )$ and $\dot{\delta} ( t - t_a )$, and by reducing the path integral (\ref{01}) with fixed end points
in configuration space to one with vanishing end points. This will permit us 
to simplify considerably the calculation of quantum mechanical correlation 
functions. To see this simplification explicitly, consider
a  harmonic oscillator whose action reads
\begin{eqnarray}
\label{03}
{\cal A} [ x ( t ) ] = \int\limits_{t_a}^{t_b} d t \, \left[
\frac{M}{2} \dot{x}^2 ( t ) - \frac{M}{2} \omega^2 x^2 ( t )  \right] \, ,
\end{eqnarray}
for which the generating functional can be calculated \cite[Eq.~(3.89)]{Kleinert} as follows:
\begin{eqnarray}
(x_b\, t_b\,|\, x_a\, t_a) [j (t)] & = &
\sqrt{\frac{M \omega}{2 \pi i \hbar \sin \omega ( t_b - t_a )}} \,
\exp \left\{ \frac{i M \omega \left[ ( x_b^2 + x_a^2 ) 
\cos \omega ( t_b - t_a ) - 2 x_a x_b 
\right]}{2 \hbar \sin \omega ( t_b - t_a )}  
\right\} \nonumber \\
&& \times \, \exp \left\{ \frac{i}{\hbar} \int\limits_{t_a}^{t_b} d t 
\, \frac{x_a
\sin\omega ( t_b - t ) + x_b \sin \omega( t - t_a )}{\sin \omega ( t_b -
t_a )}\, j ( t )  \right. \nonumber \\ 
&& \left. - \frac{i}{\hbar M \omega} \int\limits_{t_a}^{t_b} d t 
\int\limits_{t_a}^t d t' \,\frac{\sin\omega ( t_b - t ) \sin \omega ( t' 
- t_a )}{\sin \omega ( t_b - t )}\, j ( t ) j ( t' ) \right\} \, .
\label{04}
\end{eqnarray}
The nonzero end points $x_a$ and $x_b$ make this expression quite involved.
For vanishing end points, however, it simplifies to
\begin{eqnarray}
(0\, t_b\,|\, 0\, t_a) [j (t)] & = &
\sqrt{\frac{M \omega}{2 \pi i \hbar \sin \omega ( t_b - t_a )}}\nonumber \\
&& \times \, \exp \left\{
- \frac{i}{\hbar M \omega} \int\limits_{t_a}^{t_b} d t 
\int\limits_{t_a}^t d t' \,\frac{\sin\omega ( t_b - t ) \sin \omega ( t' 
- t_a )}{\sin \omega ( t_b - t )}\, j ( t ) j ( t' ) \right\} \, .
\label{04'}
\end{eqnarray}
The observation which motivates the present paper relies on 
replacing in the simple expression (\ref{04'}) the current $j ( t )$ by
\begin{eqnarray}
\label{CU}
j' ( t ) = 
j ( t ) + M x_a \dot{\delta} ( t - t_a ) - M x_b \dot{\delta} ( t_b - t ) ],
\end{eqnarray}
where the delta functions are understood as
$\dot{\delta} ( t - t_a + \epsilon )$ and $\dot{\delta} ( t_b - \epsilon - 
t )$ in the limit $\epsilon \rightarrow 0$.
By performing some partial integrations, this replacement reproduces 
all terms in the
complicated 
generating functional (\ref{04}), except for a rather trivial additional singular phase factor. The important relation is
\begin{eqnarray}
(x_b\, t_b\,|\, x_a\, t_a) [j (t)] & = &
( x_b = 0 \, t_b \,|\, x_a = 0 \, t_a ) [ j ( t ) +
M x_a \dot{\delta} ( t - t_a ) - M x_b \dot{\delta} ( t_b - t ) ]
\nonumber \\
& & \, \times \exp \left\{ \frac{i M}{2 \hbar} \, ( x_b^2 + x_a^2 ) 
\, \delta ( 0 ) 
\right\} \, .
\label{05}
\end{eqnarray}
In Section~\ref{genfunc} we prove that the relation (\ref{05}) holds for
an arbitrary quantum-mechanical system whose Hamiltonian has the
standard form
\begin{equation}
H_0 (p,x,t) = \frac{p^2}{2M} + V(x,t).
\label{14}
\end{equation}
In Section~\ref{osc} we calculate explicit amplitudes for a harmonic
oscillator with arbitrary time-dependent frequency, and as an important 
application
we derive in Section~\ref{harmfluct} from the new form of the generating 
functional a
smearing formula for calculating expectation values of
polynomial and nonpolynomial potentials functions. 
In particular, this result 
allows to calculate expectation values appearing in 
perturbation expansions for  
nonlinear interactions, as for example for the nonlinear $\sigma$-model. 
In Section V we show that our
smearing formula generalizes Wick rules and Feynman diagrams for harmonic 
expectation values from products of variables to mixtures of
nonpolynomial functions and polynomials.
In Section~\ref{genper}, we finally specialze our generating 
functional simplifies to periodic paths.  
\section{Generating Functionals}
\label{genfunc}
We begin by setting up phase space path integrals for 
generating functionals with fixed end points in either configuration
or momentum space. The action contains additional currents
$k(t)$ and $j(t)$ coupled linearily to  momentum $p ( t )$ 
and position $x ( t )$. By extending the currents with
singular $\delta$-functions as in Eq. (\ref{CU}),
we reduce the path integrals with fixed
end points to those with vanishing end points. 
Our procedure applies to arbitrary Hamiltonians $H_0 (p,x,t)$,   
with certain simplification resulting from a standard Hamiltonian (\ref{14}). 
\subsection{General Phase Space Formulation}
Consider a quantum-mechanical particle coupled to a momentum and position 
source $k ( t )$ and $j ( t )$
with the classical Hamiltonian 
\begin{equation}
H (p,x,t) = H_0 (p,x,t) - p k(t) - x j(t),
\label{1}
\end{equation}
where the corresponding action reads
\begin{equation}
A[p(t), x(t); k(t), j(t)] = \int\limits^{t_b}_{t_a} dt \Big\{
p(t) \dot x(t)  - H(p(t), x(t), t)\Big\} .
\label{2}
\end{equation}
The total time evolution amplitude between fixed space points $x_a$ and $x_b$ 
is given by the path integral
\begin{equation}
(x_b\, t_b\,|\, x_a\, t_a) [k(t), j(t)] =
\int\limits^{x_b,t_b}_{x_a,t_a} 
\frac{{\cal D}p(t) {\cal D}x(t)}{2\pi\hbar}
\exp \left\{ \frac{i}{\hbar}  {\cal A} [p(t), x(t) ; k(t), j(t)]\right\} \, .
\label{3}
\end{equation}
A Fourier transformation with respect to 
$x_a$ and $x_b$ produces the time evolution amplitude in 
momentum space
\begin{eqnarray}
\label{4}
(p_b\, t_b \,|\, p_a\, t_a) [k(t), j(t)] = 
\int\limits^{+\infty}_{-\infty} dx_a 
\int\limits^{+\infty}_{-\infty} dx_b\, e^{ -i (p_b x_b - p_a x_a)/\hbar}\,
(\,x_b\, t_b\,|\,  x_a\, t_b\,) [k(t), j(t)].
\end{eqnarray}
Here the initial and final momenta $p_a$ and $p_b$ are held fixed,
so that the right-hand side may be written as the path integral
\begin{equation}
(p_b\, t_b \,|\, p_a \, t_a) [k(t), j(t)] =
\int\limits^{p_b,t_b}_{p_a,t_a}  \frac{{\cal D}p(t)
{\cal D}x(t) }{ 2 \pi \hbar} \exp  \left\{ \frac{i}{\hbar}
{\cal A} [p(t), x(t) ; k(t), j(t)]\right\} .
\label{5}
\end{equation}
We remark that both path integrals (\ref{3}) and (\ref{5})
are properly defined as continuum limits of ordinary
integrals after a time-slicing procedure. Since end 
points of paths are fixed in coordinate and momentum space,
respectively, the discretized expressions for the path integrals
turn out to be slightly asymmetric in $p(t)$ and $x(t)$
\cite[Chap.~2]{Kleinert}.

The time evolution amplitudes (\ref{3}) and (\ref{5}) with 
fixed end points can now be reduced to corresponding ones with vanishing
end points. For this, we shift
the current $k(t)$ in (\ref{1}) by a  source term $x_{b}
\delta(t_b-t) - x_{a} \delta(t-t_a)$ and observe that this produces 
by (\ref{2}) and (\ref{5}) an overall phase factor:
\begin{eqnarray}
& & ( p_b\, t_b\,|\,  p_a\, t_a) [k(t) + x_b\delta (t_b-t) -
x_a\delta(t-t_a), j(t) ]
\nonumber \\
& & \hspace*{1cm} = \exp \left[ \frac{i}{\hbar}
( p_b x_b - p_a x_a)\right] (\,p_b\,t_b\,|\, p_a\,t_a\,)
[k(t), j(t)].
\label{8}
\end{eqnarray}
By inverting the Fourier transformation (\ref{4}), the configuration space
amplitude (\ref{3}) is seen to satisfy
\begin{eqnarray}
& & (x_b \, t_b\,|\, x_a\, t_a) [k(t) + 
{x'}_{\hspace*{-0.1cm}b} \hspace*{0.1cm} 
\delta (t_b -t) - {x'}_{\hspace*{-0.1cm}a} \hspace*{0.1cm}
\delta (t-t_a), j(t)] \nonumber \\
& & \hspace*{1cm} = (\,x_b +  {x'}_{\hspace*{-0.1cm}b} \hspace*{0.1cm}
\,t_b \,|\, x_a + {x'}_{\hspace*{-0.1cm}a} \hspace*{0.1cm}
\,t_a\,)
[k(t), j(t)]\, , 
\label{9}
\end{eqnarray}
where again the delta functions are understood as
$\delta (t_b - \epsilon - t)$ and $\delta ( t - t_a + \epsilon )$
in the limit $\epsilon \rightarrow 0$.
Because of this relation, the amplitude (\ref{3}) can be reduced to a 
path integral with
vanishing end points but additional
$\delta$-terms in the current $k(t)$:
\begin{equation}
(x_b\, t_b \,|\, x_a\, t_a) [k(t), j(t) ] = (x_b = 0\, t_b\,|\,
x_a = 0\, t_a) [k(t) + x_b \delta (t_b - t) - x_a \delta (t-t_a),j(t)].
\label{10}
\end{equation}
A similar expression exists, if momentum end points are fixed in momentum space by adding
$p_a \delta (t-t_a) - p_b \delta (t_b - t)$ to the current $j(t)$:
\begin{equation}
(p_b\,t_b \,|\, p_a\, t_a) [k(t),j(t)] =  (p_b = 0\, t_b\,|\,
p_a = 0\,t_a) [k(t), j(t) + p_a \delta (t-t_a) - p_b\delta
(t_b-t)].
\label{11}
\end{equation}
We now explore the 
consequences of these two relations 
for the calculation of correlation functions.
\subsection{Correlation Functions}
\label{corrfunc}
The functional dependence of the time evolution
amplitudes (\ref{3}) and (\ref{5}) on the currents $k(t)$
and $j(t)$ allows us to calculate expectation values of arbitrary 
functionals $F[p(t), x(t)]$ from the path integral 
\begin{eqnarray}
& & \langle F[p(t), x(t)]\rangle  [k(t), j(t)]^{v_b,t_b}_{v_a,t_a} 
= \frac{1}{(v_b \, t_b\,|\, v_a \, t_a)[k(t),j(t)]}
\nonumber \\ 
& & \hspace*{1cm} \times \int\limits^{v_b,t_b} _{v_a,t_a}
\frac{{\cal D} p(t) {\cal D}x(t)}{2\pi \hbar} F[p(t),x(t)]\exp
\left\{ \frac{i}{\hbar}{\cal A}[p(t),x(t);k(t),j(t)]\right\}  
\label{6} \, ,
\end{eqnarray}
where the variable $v$ may be $p$ or $x$.
The usual correlation functions 
\begin{eqnarray}
& & \langle p ( t_1 ) \cdots p ( t_n )  x ( t_1 ) \cdots 
x ( t_m ) \rangle  [k(t), j(t)]^{v_b,t_b}_{v_a,t_a} 
= \frac{1}{(v_b \, t_b\,|\, v_a \, t_a)[k(t),j(t)]}
\nonumber \\ 
& & \hspace*{1cm} \times \int\limits^{v_b,t_b} _{v_a,t_a}
\frac{{\cal D} p(t) {\cal D}x(t)}{2\pi \hbar}
p ( t_1 )\cdots p ( t_n )  x ( t_1 )  \cdots 
x ( t_m )  \exp
\left\{ \frac{i}{\hbar}{\cal A}[p(t),x(t);k(t),j(t)]\right\}  
\label{6a} \, ,
\end{eqnarray}
are special cases of (\ref{6}), so we shall call the
general expectation values (\ref{6}) {\it correlation functionals}. The sources
$k ( t )$ and $j ( t )$ permit us to express (\ref{6}) in terms of
functional derivatives:
\begin{equation}
\langle F[p(t) , x(t)]\rangle [k(t), j(t)]^{v_b,t_b}_{v_a,t_a}=
\frac{{\displaystyle F\left[\frac{\hbar}{i}\frac{\delta }{\delta k(t)},
\frac{\hbar}{i} \frac{\delta}{\delta j (t)}\right] 
(v_b\,t_b\,| \, v_a\,t_a ) [k(t),j(t)] }}
{(v_b \, t_b\, |\, v_a \, t_a)[k(t),j(t)]}.
\label{7}
\end{equation}
Recalling (\ref{10}) and (\ref{11}), we shall rewrite the functionals $(v_b\,t_b\,| \, v_a\,t_a ) [k(t),j(t)]$
in a unified common way as follows
\begin{eqnarray}
& & (v_b\,t_b \,|\, v_a\, t_a) [k(t), j(t) ] \nonumber \\
& & \hspace*{0.8cm} = \int\limits^{w_b=0,t_b}_{w_a=0,t_a}  
\frac{{\cal D} p(t) {\cal D}x(t)}{2\pi\hbar}
\delta (v(t_a)  - v_a) \delta (v (t_b)-v_b) \exp \left\{ \frac{i}{\hbar}
{\cal A} [ p(t), x(t); k(t), j(t)]\right\}
\label{12} \, ,
\end{eqnarray}
where the paths $v ( t ) $ stand either for $p ( t )$ or for $x ( t )$. In each of these cases, the paths $w ( t )$ denote the conjugate variables $x ( t )$ or $p ( t )$, respectively. 
In this form, the path integral possesses the advantage
that usual correlation functions (\ref{6a}) can be
determined by path averages, in which intermediate
and end points are treated on equal footing. Indeed, inserting delta functions
according to 
\begin{eqnarray}
\nonumber 
& & \langle p(t_1) \cdots  p(t_n)  x (t_{1}) \cdots 
x(t_m) \rangle [k(t), j(t)] ^{v_b, t_b}_{v_a,t_a} \\
&  & \hspace*{0.8cm} =\frac{1}{(v_b \, t_b\,|\, v_a\, t_a) [k(t), j(t)] }
\int\limits^{+\infty}_{-\infty}d p_1 \cdots
\int\limits^{+\infty}_{-\infty} d p_n \int\limits^{+\infty}_{-\infty}
dx_{1} \cdots \int\limits^{+\infty}_{-\infty} dx_m\,
p_1 \cdots p_n x_{1} \cdots  x_m \nonumber \\
& &  \hspace*{1.3cm} \times \, \int\limits^{v_b,t_b}_{v_a,t_a} 
\frac{{\cal D} p(t) {\cal D}x (t)}
{2 \pi \hbar} \delta  (p(t_1)-
p_1) \cdots  \delta (p(t_n) - p_n)  
\nonumber \\
& & \hspace*{1.3cm}\times \delta (x(t_{1}) - x_{1})
\cdots \delta
(x (t_m) - x_m) \exp \left\{ \frac{i}{\hbar} {\cal A}
[p(t), x(t); k(t), j(t) ]\right\} \, ,
\label{13a}
\end{eqnarray}
we obtain with a similar reasoning
\begin{eqnarray}
\nonumber 
& & \langle p(t_1) \cdots  p(t_n)  x (t_{1}) \cdots 
x(t_m) \rangle [k(t), j(t)] ^{v_b, t_b}_{v_a,t_a} \\
&  & \hspace*{0.8cm} =\frac{1}{(v_b \, t_b\,|\, v_a\, t_a) [k(t), j(t)] }
\int\limits^{+\infty}_{-\infty}d p_1 \cdots
\int\limits^{+\infty}_{-\infty} d p_n \int\limits^{+\infty}_{-\infty}
dx_{1} \cdots \int\limits^{+\infty}_{-\infty} dx_m\,
p_1 \cdots p_n x_{1} \cdots  x_m \nonumber \\
& &  \hspace*{1.3cm} \times \, \int\limits^{w_b=0,t_b}_{w_a=0,t_a} 
\frac{{\cal D} p(t) {\cal D}x (t)}
{2 \pi \hbar} \delta (v (t_a) - v_a)   \delta  (p(t_1)-
p_1) \cdots  \delta (p(t_n) - p_n)  
\nonumber \\
& & \hspace*{1.3cm}\times \delta (x(t_{1}) - x_{1})
\cdots \delta
(x (t_m) - x_m)
\delta (v (t_b)  - v_b )\exp \left\{ \frac{i}{\hbar} {\cal A}
[p(t), x(t); k(t), j(t) ]\right\} \, .
\label{13}
\end{eqnarray}
\subsection{Standard Hamiltonian}
The above formalism can be made more specific for
the standard Hamiltonian (\ref{14}).
Then the path integrals over the momentum paths  $p(t)$ in (\ref{3})
and (\ref{5}) becomes harmonic and can be explicitly evaluated.
The phase space integral (\ref{3}), for instance,  
reduces to the configuration space path integral
\begin{eqnarray}
(x_b\, t_b\,|\, x_a\, t_a) [k(t), j(t)]
= \int\limits_{x_a,t_a}^{x_b,t_b}
 {\cal D} x (t) \exp \left\{
\frac{i}{\hbar}  {\cal A} [ x(t) ; k ( t ) , j(t) ] \right\} \, ,
\label{15}
\end{eqnarray}
where the current $k(t)$ couples linearily to the path momentum
$M \dot x (t)$ in the action 
\begin{equation}
{\cal A} [x(t); k ( t ) , j (t)] = 
\int\limits^{t_b}_{t_a} dt \left\{ \frac{M}{2}
\dot{x}^2 (t) - V (x(t),t) + x (t) j(t)
+ M \dot x (t) k(t) +  \frac{M}{2} k^2(t)  \right\} .
\label{16}
\end{equation}
A subsequent partial integration transforms
the current $k(t)$ to an effective coordinate current
with an extra phase factor:
\begin{eqnarray}
\nonumber
& & (x_b\, t_b\,|\, x_a\, t_a) [k(t), j(t)] \\
& & \hspace*{1cm} = ( x_b\, t_b\,|\,
x_a\,t_a) [0, j(t) - M\dot k (t)] \exp \left\{
\frac{iM}{\hbar} \left[ x_b k_b - x_a k_a + \frac{1}{2}
\int\limits^{t_b}_{t_a}  dt \, k^2 (t)\right] \right\} \, . 
\label{17}
\end{eqnarray}
In the next section
we determine the generating functional 
$(x_b\, t_b\,|\, x_a\, t_a) [ 0 , j(t)]$ for a harmonic oscillator
with arbitrary time-dependent frequency $\Omega ( t )$ and use
(\ref{17})
to construct the full generating functional
$(x_b\, t_b\,|\, x_a\, t_a) [k(t), j(t)]$.
\section{Time-Dependent Harmonic Oscillator}
\label{osc}
Consider a standard Hamiltonian (\ref{14}) with
a harmonic potential containing an arbitrary time-dependent frequency:
\begin{eqnarray}
\label{18}
V ( x , t ) = \frac{M}{2} \Omega^2 ( t ) x^2 \, .
\end{eqnarray}
The generating functionals (\ref{3})
and (\ref{5}) are then expressable in terms of two 
fundamental solutions $D_a ( t ), D_b ( t )$ of the corresponding
classical equation of motion with particular boundary conditions
\cite{Kleinert1}
\begin{eqnarray}
\label{19}
\hat{K} ( t ) \, D_a ( t ) = 0 \, ;& 
\hspace*{1cm} & D_a ( t_a ) = 0 \, , \, \dot{D}_a ( t_a ) = 1 \, , \\
\label{20}
\hat{K} ( t ) \,  D_b ( t ) = 0 \, ; 
& \hspace*{1cm} & D_b ( t_b ) = 0 \, , \, \dot{D}_b ( t_b ) = - 1 \, ,
\end{eqnarray}
where $\hat{K} ( t )$ denotes the operator
\begin{eqnarray}
\label{20B}
\hat{K} ( t ) = - \partial_t^2 - \Omega^2 ( t ) \, .
\end{eqnarray}
Since the
time derivative of the Wronski determinant
\begin{eqnarray}
\label{21}
W ( t ) = D_a ( t ) \dot{D}_b ( t ) - \dot{D}_a ( t ) D_b ( t )
\end{eqnarray}
vanishes, we observe the identity
\begin{eqnarray}
\label{22}
D_a ( t_b ) = D_b ( t_a ) \,.
\end{eqnarray}
Note that a similar identity does not hold for the time derivatives of
the two fundamental solutions $D_a ( t )$ and $D_b ( t )$. Indeed, 
partially integrating the differential equation for $\dot{D}_a ( t )$
and taking into account (\ref{19})-(\ref{20}), we deduce
\begin{eqnarray}
\dot{D}_b ( t_a ) + \dot{D}_a ( t_b ) = - 2 \int\limits_{t_a}^{t_b}
d t \, \Omega ( t ) \dot{\Omega} ( t ) \, D_a ( t ) D_b ( t ) \, .
\end{eqnarray}
Let us now determine the time evolution amplitude (\ref{15})
in configuration space
for a vanishing current $k ( t )$ as defined in (\ref{15}).
We decompose the paths $x ( t )$ into the classical path $x_{\rm cl}^j ( t )$
and the quantum fluctuations $\delta x ( t )$ around it:
\begin{eqnarray}
\label{25}
x ( t ) = x_{\rm cl}^j ( t ) + \delta x ( t ) \, .
\end{eqnarray}
The classical path $x_{\rm cl}^j ( t )$ solves the boundary value problem
\begin{eqnarray}
\label{23}
\hat{K} ( t ) \, x_{\rm cl}^j ( t ) = - 
\frac{j ( t )}{M} \, ; \hspace*{1cm} x_{\rm cl}^j ( t_a ) = x_a \, , \, 
x_{\rm cl}^j ( t_b ) = x_b  \, ,
\end{eqnarray}
and the fluctuations $\delta x ( t )$ vanish at the endpoints: 
\begin{eqnarray}
\label{24}
\delta x ( t_a ) = \delta x ( t_b ) = 0 \, .
\end{eqnarray}
Inserting the decomposition (\ref{25}) into the action 
(\ref{16}), we observe that due
to (\ref{25}) and (\ref{23}) the total action decomposes into a classical part
\begin{eqnarray}
\label{26}
{\cal A} [ x_{\rm cl}^j ( t ) ; 0 , j ( t ) ] = \frac{M}{2} \left[ x_b 
\dot{x}_{\rm cl}^j ( t_b ) - x_a \dot{x}_{\rm cl}^j ( t_a ) 
\right] + \frac{1}{2} \int\limits_{t_a}^{t_b} d t \, x_{\rm cl}^j ( t )
j ( t ),
\end{eqnarray}
and a fluctuation part, which is simply the classical action 
evaluated for the fluctuations $\delta x ( t )$ at $j = 0$:
\begin{eqnarray}
\label{27}
{\cal A} [ x ( t ) ; 0 , j ( t ) ] = {\cal A} [ x_{\rm cl}^j ( t ) ; 0 , 
j ( t ) ] + {\cal A} [ \delta x ( t ) ; 0 , 0 ] \, .
\end{eqnarray}
Inserting this into
the original path integral (\ref{15}), it 
factorizes into the product of a classical amplitude 
with the classical
action (\ref{26}), and an additional fluctuation factor which is equal to the
amplitude at vanishing end points:
\begin{eqnarray}
\label{28}
(\, x_b \, t_b \,|\, x_a \, t_a \,) [ 0 ; j ( t ) ] = \exp
\left\{ \frac{i}{\hbar} {\cal A} [ x_{\rm cl}^j ( t ) ; 0 , j ( t ) ] 
\right\} ( \,x_b = 0 \, t_b \,|\, x_a = 0 \, t_a\, ) [ 0 , 0 ] \, .
\end{eqnarray}
\subsection{Classical Action}

The classical action in the presence of currents can be expressed in terms
of the solutions $D_a ( t ), D_b ( t )$ of the time-dependent harmonic
boundary value problems
(\ref{19}) and (\ref{20}). 
First we decompose the solution 
of the boundary value problem (\ref{23}) in the presence of 
into a homogeneous
and an inhomogeneous contribution:
\begin{eqnarray}
\label{29}
x_{\rm cl}^j ( t ) = 
x_{\rm cl} ( t ) +
\Delta x_{\rm cl}^j ( t )  \, .
\end{eqnarray}
The homogeneous solution reads
\begin{eqnarray}
\label{30}
x_{\rm cl} ( t ) = \frac{D_b ( t ) x_a 
+ D_a ( t ) x_b}{D_a ( t_b )} \, ,
\end{eqnarray}
while the inhomogeneous one is given by
\begin{eqnarray}
\label{31}
\Delta x_{\rm cl}^j ( t )  = - \frac{1}{M} \int\limits_{t_a}^{t_b}
d t' \, G_{jj}^x ( t , t' ) j ( t' ) \, ,
\end{eqnarray}
where $G_{jj}^x ( t , t' )$ denotes the Green function of the classical
equation of motion
\begin{eqnarray}
\label{32}
\hat{K} ( t ) \, G_{jj}^x ( t , t' ) = 
\delta ( t - t' ), 
\end{eqnarray}
with Dirichlet boundary conditions
\begin{eqnarray}
\label{33}
G_{jj}^x ( t_a , t' ) = G_{jj}^x ( t_b , t' ) = 0 \, .
\end{eqnarray}
>From (\ref{32}) we deduce that the Green function $G_{jj}^x ( t , t' )$ solves
the homogeneous differential equation for $t \neq t'$:
\begin{eqnarray}
\label{34}
\hat{K} ( t )  G_{jj}^x ( t , t' ) = 0 \, ,
\end{eqnarray}
and that its first derivative $\partial_t G_{jj}^x ( t , t' )$ is 
discontinuous at $t = t'$:
\begin{eqnarray}
\label{35}
\lim_{\epsilon \downarrow 0} \left[ \left. \partial_t 
G_{jj}^x ( t , t' ) \right|_{t = t' + \epsilon} -
\left. \partial_t G_{jj}^x ( t , t' ) \right|_{t = t' - \epsilon} \right]
= -1 \, .
\end{eqnarray}
The Green function itself
is continuous around $t = t'$:
\begin{eqnarray}
\label{36}
\lim_{\epsilon \downarrow 0} \left[ 
\left. G_{jj}^x ( t , t' ) \right|_{t = t' + \epsilon} -
\left. G_{jj}^x ( t , t' ) \right|_{t = t' - \epsilon} \right]
= 0 \, .
\end{eqnarray}
The solution of 
(\ref{33})-(\ref{36}) is given by Wronski's famous expression
\begin{eqnarray}
\label{37}
G_{jj}^x ( t , t' ) = \frac{\Theta ( t - t' ) D_b ( t ) D_a ( t' ) + 
\Theta ( t' - t )   D_a ( t ) D_b ( t' )}{D_a ( t_b )} =
G_{jj}^x ( t' , t ) \, ,
\end{eqnarray}
where $\Theta ( t - t' ) $ 
denotes the Heaviside function which vanishes for $t 
< t'$ and is equal to unity for $t > t'$. 
Inserting (\ref{29}) and (\ref{31}) we obtain for the 
classical action (\ref{26})
\begin{eqnarray}
{\cal A} [ x_{\rm cl}^j ( t ) ; 0  , j ( t ) ] &=& \frac{M}{2 D_a ( t_b )}
\left[ \dot{D}_a ( t_b ) x^2_b - \dot{D}_b ( t_a ) x^2_a - 2
x_a x_b \right] \nonumber \\
&& + \int\limits_{t_a}^{t_b} d t \, x_{\rm cl} ( t ) 
j ( t ) - \frac{1}{2 M} \int\limits_{t_a}^{t_b} d t 
\int\limits_{t_a}^{t_b} d t' \, G_{jj}^x ( t , t' ) j ( t ) j ( t' ) \, ,
\label{38}
\end{eqnarray}
where $x_{\rm cl} ( t )$ and $G_{jj}^x ( t , t' )$
are given by (\ref{30}) and (\ref{37}), respectively.
\subsection{Fluctuation Factor}
Now we calculate the fluctuation factor in (\ref{28}). 
Recalling the path representation (\ref{15}) with the action (\ref{16}), we have to evaluate
\begin{eqnarray}
\label{40}
(x_b= 0\, t_b\,|\, x_a= 0\, t_a) [0,0] = 
\int\limits_{\delta x_a= 0,t_a}^{\delta x_b = 0,t_b}
{\cal D} \delta x (t) \exp \left[
\frac{iM }{2 \hbar} \int\limits^{t_b}_{t_a} dt \, \delta x ( t ) 
\hat{K} ( t ) \, \delta x ( t ) \right] \, . 
\end{eqnarray}
To this end we decompose the fluctuations $\delta x ( t )$ in 
(\ref{40}) into eigenfunctions $x_n ( t )$ of the operator $\hat{K}(t)$ of (\ref{20B}) with Dirichlet boundary conditions
\begin{eqnarray}
\label{41}
\hat{K} ( t )  \, x_n ( t )  =  \lambda_n \, x_n ( t )\, ;
\hspace*{1cm} x_n ( t_a ) = x_n ( t_b ) = 0 
\end{eqnarray}
which satisfy the orthonormality and completeness relations
\begin{eqnarray}
\int\limits_{t_a}^{t_b} d t \, x_{n} (t ) \, x_{n'} (t ) & = & 
\delta_{n,n'} \, , \label{42}\\
\sum_n x_n ( t ) \, x_n (t' ) & = & \delta ( t - t' ) \, , \label{43}
\end{eqnarray}
as follows:
\begin{eqnarray}
\label{44}
\delta x ( t ) = \sum_n c_n \, x_n ( t ) \, .
\end{eqnarray}
The path integral over all possible fluctuations $\delta x ( t )$ in 
(\ref{40}) amounts to a product of integrals over all expansion 
coefficients $c_n$: 
\begin{eqnarray}
\label{45}
\int\limits_{\delta x ( t_a ) = 0}^{\delta x ( t_b ) = 0}
{\cal D} \delta x (t) = J \, \left\{ \prod_n \int\limits_{- \infty}^{+
\infty} d c_n \right\} \, .
\end{eqnarray}
The Jacobi determinant $J$ of the transformation (\ref{44}) is an irrelevant
constant.
Applying (\ref{41})-(\ref{45}), the path integral (\ref{40}) is finally
determined by 
\begin{eqnarray}
\label{46}
(x_b= 0\, t_b\,|\, x_a= 0\, t_a) [0,0] = \frac{J}{\sqrt{\mbox{Det}
\hat{K} ( t )}} \, , 
\end{eqnarray}
where the determinant of the operator $\hat{K} ( t )$ is equal to the 
product of its eigenvalues
\begin{eqnarray}
\label{47}
\mbox{Det} \, \hat{K} ( t ) = \prod_n \lambda_n.
\end{eqnarray}
\subsection{Operator Determinant}
In order to calculate the operator determinant (\ref{47})
it is advantageous to introduce a one-parameter family of 
operators \cite{Chervyakov1,Chervyakov2} 
\begin{eqnarray} 
\label{39}
\hat{K}^g ( t ) = - \partial_t^2 - g\,\Omega^2 ( t )  \, ,
\end{eqnarray}
depending linearily on a coupling strength parameter 
$g \in [ 0 , 1 ]$, and coinciding with the original
operator $\hat{K} ( t )$ in (\ref{20B})
for $g = 1$. It is possible to
relate the operator determinant $\mbox{Det} \, \hat{K}^{g} ( t )$
to the fundamental solutions
$D^g_a ( t )$, $D^g_b ( t )$, 
and to the Green
function $G^{x,g}_{jj} ( t , t' )$ emerging from (\ref{19}), (\ref{20})
and (\ref{32}), (\ref{33}).
For this we substitute the operator $\hat{K} ( t )$ by
$\hat{K}^g ( t )$, and differentiate the $g$-dependent version of the eigenvalue problem (\ref{41}) with respect to $g$:
\begin{eqnarray}
\label{49}
\hat{K}^g ( t ) \, \frac{\partial x_n^g ( t )}{\partial g} - \Omega^2
( t ) \, x_n^g ( t ) = \frac{\partial \lambda_n^g}{\partial g}
\, x_n^g ( t ) + \lambda_n^g\, \frac{\partial x_n^g ( t )}{\partial g} \, .
\end{eqnarray}
Multiplying (\ref{49}) with $x_n^g ( t ) / \lambda_n^g$ and
performing a summation over $n$ plus an integration with respect to $t$, we
obtain with (\ref{41}), (\ref{42}) and (\ref{47}):
\begin{eqnarray}
\label{50}
\frac{\partial }{\partial g} \ln \mbox{Det} \, \hat{K}^g ( t ) = - 
\int\limits_{t_a}^{t_b} d t \, \Omega^2 ( t ) \, G^{x,g}_{jj} ( t , t )
\, .
\end{eqnarray}
In the last step
we have used the spectral decomposition of the Green function
\begin{eqnarray}
\label{52}
G^{x,g}_{jj} ( t , t' ) = 
\sum_n \frac{x_n^g ( t ) \, x_n^g ( t' )}{\lambda_n^g} \, .
\end{eqnarray} 
To solve the differential equation (\ref{50}),
we differentiate the boundary value equation (\ref{19}) for
$D^g_a ( t )$ with respect to $g$, and obtain the inhomogeneous
initial value problem
\begin{eqnarray}
\label{53}
\hat{K}^g ( t ) \frac{\partial D^g_a ( t )}{\partial g} = \Omega^2 ( t ) \,
D^g_a ( t ) \, ; \hspace*{1cm} \left. \frac{\partial D^g_a ( t )}{\partial g}
\right|_{t = t_a} = \left. \frac{\partial}{\partial t} \, \frac{\partial
D^g_a ( t )}{\partial g} \right|_{t = t_a} = 0 \, .
\end{eqnarray}
Generalizing (\ref{37}) from $g = 1$ to arbitrary values $g \in [ 0 , 1 ]$, the solution of (\ref{53}) is given by
\begin{eqnarray}
\label{54}
\frac{\partial}{\partial g} \, \ln D^g_a ( t_b ) = - 
\int\limits_{t_a}^{t_b} d t \, \Omega^2 ( t ) \, G^{x,g}_{jj} ( t , t ) 
\, .
\end{eqnarray}
This shows that (\ref{50}) is solved by
\begin{eqnarray}
\mbox{Det} \, \hat{K}^g ( t ) = C \,  D^g_a ( t_b ) \, ,
\end{eqnarray}
where $C$ denotes some constant. Due to this result, the ratio of two
fluctuation factors with two different parameters $g_1$ and $g_2$ can be rewritten as
\begin{eqnarray}
\label{55}
\frac{(x_b= 0\, t_b\,|\, x_a= 0\, t_a) 
[0,0]^{g_1}}{(x_b= 0\, t_b\,|\, x_a= 0\, t_a) [0,0]^{g_2}} = 
\sqrt{\frac{D^{g_2}_a ( t_b )}{D^{g_1}_a ( t_b )}}
\, .
\end{eqnarray}
This serves to determine the fluctuation factor of the initial
time-dependent harmonic oscillator at $g_1 = 1$ in terms of
the fluctuation factor of the free particle $g_2 = 0$. The latter is well-known
and may be calculated explicitly, for instance, via time-slicing
\cite[Chap.~2]{Kleinert} as
\begin{eqnarray}
(\,x_b= 0\, t_b\,|\, x_a= 0\, t_a\,) [0,0]^{g_2 = 0} = 
\sqrt{\frac{M}{2 \pi i \hbar (t_b - t_a )}} \, .
\end{eqnarray}
Since the obvious solution of (\ref{19}) at $g_2 = 0$ reads
$D_a^{g_2=0} ( t_b )= t_b - t_a$,
we obtain the famous
Gelfand-Yaglom formula for Dirichlet boundary conditions \cite{GY}:
\begin{eqnarray}
\label{56}
(\,x_b= 0\, t_b\,|\, x_a= 0\, t_a\,) [0,0]
= \sqrt{\frac{M}{2 \pi i \hbar D_a ( t_b )}} \, .
\end{eqnarray}
Note that similar results 
can also be derived for periodic and antiperiodic
boundary conditions \cite{Chervyakov1,Chervyakov2}.
\subsection{Full Generating Functional}
Having obtained
the generating functional $(x_b\, t_b\,|\, x_a\, t_a) [0, j(t)]$ 
of the harmonic oscillator with arbitrary frequency with 
vanishing current $k ( t )$, we now make use of the relation
(\ref{17}) to derive the full
generating functional $(x_b\, t_b\,|\, x_a\, t_a) [k(t), j(t)]$.  
The terms containing 
the current velocity $\dot{k} ( t )$ can be turned into functionals of 
$k ( t )$ itself with the help of
several partial integrations. These turn out to
remove the extra phase factor in (\ref{17}). As a
result, the time evolution amplitude in the configuration 
representation is determined by a Van Vleck-Pauli-Morette type of formula
\cite[Chap.~4]{Kleinert}
\begin{eqnarray}
&&(x_b\, t_b\,|\, x_a\, t_a) [k(t), j(t)] \nonumber \\
& & \hspace*{1cm} = \sqrt{\frac{i}{2 \pi \hbar} 
\, \frac{\partial^2 {\cal A} ( x_b , t_b ; x_a , t_a ) [ k ( t ) , 
j ( t )]}{\partial x_b \partial x_a}} 
\label{57} 
\exp \left\{ \frac{i}{\hbar}
{\cal A} ( x_b , t_b ; x_a , t_a ) [ k ( t ) , 
j ( t )] \right\}
\end{eqnarray}
with the action
\begin{eqnarray}
& & {\cal A} ( x_b , t_b ; x_a , t_a ) [ k ( t ) , 
j ( t )] = \frac{M \left[\dot{D}_a ( t_b ) x^2_b - 
\dot{D}_b ( t_a ) x^2_a - 2
x_a x_b \right]}{2 D_a ( t_b )} 
\nonumber \\
& & \hspace*{1cm} +\int\limits_{t_a}^{t_b} d t \,\left[
x_{\rm cl} ( t ) j ( t ) 
+ p_{\rm cl} ( t ) k ( t ) \right] 
- \frac{1}{2}
\int\limits_{t_a}^{t_b} d t \int\limits_{t_a}^{t_b} d t' \Bigg[
\frac{1}{M} G_{jj}^x ( t , t' ) j ( t ) j ( t' ) \nonumber \\
& & \hspace*{1cm} + G_{jk}^x ( t , t' ) j ( t ) k ( t' )
+ G_{kj}^x ( t , t' ) k ( t ) j ( t' )
+ M G_{kk}^x ( t , t' ) k ( t ) k ( t' ) \Bigg] \, .
\label{57b}
\end{eqnarray}
The homogeneous classical solution $x_{\rm cl} ( t )$ is given in
(\ref{30}), and $p_{\rm cl} ( t )$ denotes the classical momentum
$p_{\rm cl} ( t )\equiv M
\dot{x}_{\rm cl} ( t )$. The Green function $G_{jj}^x ( t , t' )$ is
given by (\ref{37}), while the others are 
\begin{eqnarray}
\label{58}
G_{jk}^x ( t , t' ) & = & \frac{\Theta ( t - t' ) D_b ( t ) \dot{D}_a ( t' ) + 
\Theta ( t' - t )   D_a ( t ) \dot{D}_b ( t' )}{D_a ( t_b )} =
G_{kj}^x ( t' , t ) \, , \\
\label{59}
G_{kk}^x ( t , t' ) & = & \frac{\Theta ( t - t' ) \dot{D}_b ( t ) 
\dot{D}_a ( t' ) + \Theta ( t' - t )  \dot{D}_a ( t ) 
\dot{D}_b ( t' )}{D_a ( t_b )} =
G_{kk}^x ( t' , t ) \, . 
\end{eqnarray}
By differentiating (\ref{57b}) functionally with respect to $j$ and $k$, we see that the Green functions correspond to the correlation functions
\begin{eqnarray}
  \label{59a}
  \langle\, x_b\,|\,\tx(t)\,\tx(t')\,|\,x_a\,\rangle &=&\frac{i\hbar}{M}G_{jj}^x(t,t'),\\
  \label{59b}
  \langle\, x_b\,|\,\tx(t)\,\tp(t')\,|\,x_a\,\rangle &=&i\hbar G_{jk}^x(t,t')=i\hbar G_{kj}^x(t',t),\\
  \label{59c}
  \langle\, x_b\,|\,\tp(t)\,\tp(t')\,|\,x_a\,\rangle &=&i\hbar MG_{kk}^x(t,t')
\end{eqnarray}
with $\tx(t)=x(t)-x_{\rm cl}(t)$ and $\tp(t)=p(t)-p_{\rm cl}(t)$.
These results can be summarized by the mnemonic rule that the Green functions
involving a momentum current $k ( t )$ once or twice follow from 
$G_{jj}^x ( t , t' )$ by one or two time derivatives if the
time derivatives of the Heaviside functions are neglected:
\begin{eqnarray}
G_{jk}^x ( t , t' ) = \frac{\partial G_{jj}^x ( t , t' )}{\partial t}\, ,
\hspace*{0.3cm}
G_{kj}^x ( t , t' ) = \frac{\partial G_{jj}^x ( t , t' )}{\partial t'}\, ,
\hspace*{0.3cm}
G_{kk}^x ( t , t' ) = \frac{\partial^2 G_{jj}^x ( t , t' )}{\partial t.
\partial t'}\, ,
\hspace*{0.3cm}
\end{eqnarray}
A complete analogous expression to (\ref{57b}) is found for
the time evolution amplitude in the momentum representation.
The Fourier transformation (\ref{4}) of (\ref{57}) yields
a Van Vleck-Pauli-Morette type of formula 
\begin{eqnarray}
&&(p_b\, t_b\,|\, p_a\, t_a) [k(t), j(t)] \nonumber \\
& & \hspace*{1cm} = \sqrt{2 \pi i \hbar 
\, \frac{\partial^2 {\cal A} ( p_b , t_b ; p_a , t_a ) [ k ( t ) , 
j ( t )]}{\partial p_b \partial p_a}} 
\label{60b}
\exp \left\{ \frac{i}{\hbar}
{\cal A} ( p_b , t_b ; p_a , t_a ) [ k ( t ) , 
j ( t )] \right\}
\end{eqnarray}
where the action is the Legendre transform of  (\ref{57b})
\begin{eqnarray}
\label{LT}
{\cal A} ( p_b , t_b ; p_a , t_a ) [ k ( t ) , j ( t )] = 
{\cal A} ( x_b , t_b ; x_a , t_a ) [ k ( t ) , j ( t )]
- p_b x_b + p_a x_a,
\end{eqnarray}
calculated for the conjugate variables
\begin{eqnarray}
p_b = \frac{\partial {\cal A} ( x_b , t_b ; x_a , t_a ) 
[ k ( t ) , j ( t )]}{\partial x_b} \, , \hspace*{1cm}
p_a = - \frac{\partial {\cal A} ( x_b , t_b ; x_a , t_a ) 
[ k ( t ) , j ( t )]}{\partial x_a} \, .
\end{eqnarray}
This brings (\ref{LT}) to the form
\begin{eqnarray}
&& {\cal A} ( p_b , t_b ; p_a , t_a ) [ k ( t ) , j ( t )]  = 
\frac{D_a ( t_b ) \left[
\dot{D}_a ( t_b ) p^2_a - \dot{D}_b ( t_a ) p^2_b - 2
p_a p_b \right]}{2 M [ 1 + \dot{D}_a ( t_b ) \dot{D}_b ( t_a )]} 
\nonumber \\
& & \hspace*{1cm} + \int\limits_{t_a}^{t_b} d t \,\left[
\bar{x}_{\rm cl} ( t ) j ( t )
+ \bar{p}_{\rm cl} ( t ) k ( t ) \right] - \frac{1}{2}
\int\limits_{t_a}^{t_b} d t \int\limits_{t_a}^{t_b} d t' \Bigg[
\frac{1}{M} G_{jj}^p ( t , t' ) j ( t ) j ( t' ) 
\nonumber\\
& & \hspace*{1cm} + G_{jk}^p ( t , t' ) j ( t ) k ( t' )
+ G_{kj}^p ( t , t' ) k ( t ) j ( t' )
+ M G_{kk}^p ( t , t' ) k ( t ) k ( t' ) \Bigg] \, ,
\label{61}
\end{eqnarray}
where the classical solution now reads
\begin{eqnarray}
\label{62}
\bar{x}_{\rm cl} ( t ) = \frac{p_a [ D_a ( t ) + 
D_b ( t ) \dot{D}_a ( t_b )] + p_b [ D_a ( t ) \dot{D}_b ( t_a ) - 
D_b ( t )]}{M [ 1 + \dot{D}_a( t_b ) \dot{D}_b ( t_a ) ]} \, ,
\end{eqnarray}
and $\bar{p}_{\rm cl} ( t )$ denotes the associated classical 
momentum $\bar{p}_{\rm cl} ( t ) \equiv M
\dot{\bar{x}}_{\rm cl} ( t )$. The Green functions in (\ref{61})
turn out to be
\begin{eqnarray}
G_{jj}^p ( t , t' ) & = & 
\Theta ( t - t' ) \frac{[ D_b ( t ) 
\dot{D}_a ( t_b ) + D_a ( t )] [ D_a ( t' ) \dot{D}_b ( t_a ) - 
D_b ( t' )]}{D_a ( t_b )[1 + 
\dot{D}_a( t_b ) \dot{D}_b ( t_a )]} \nonumber \\
&& + \Theta ( t' - t ) 
\frac{[ D_a ( t ) \dot{D}_b ( t_a ) - 
D_b ( t )] [ D_b ( t' ) \dot{D}_a ( t_b )+ D_a ( t' )]}{D_a ( t_b )[1 + 
\dot{D}_a( t_b ) \dot{D}_b ( t_a )]} 
= G_{jj}^p ( t' , t ) \, ,
\label{63}\\
G_{jk}^p ( t , t' ) & = &  \Theta ( t - t' ) \frac{[ D_b ( t )  
\dot{D}_a ( t_b ) + D_a ( t )] [ \dot{D}_a ( t' ) \dot{D}_b ( t_a ) - 
\dot{D}_b ( t' )]}{D_a ( t_b )[1 + 
\dot{D}_a( t_b ) \dot{D}_b ( t_a )]} \nonumber \\ 
& & +  \Theta ( t' - t ) \frac{
[ D_a ( t ) \dot{D}_b ( t_a ) - 
D_b ( t )] [ \dot{D}_{b} ( t' )  \dot{D}_a ( t_b ) 
+ \dot{D}_a ( t' )]}{D_a ( t_b )[1 + 
\dot{D}_a( t_b ) \dot{D}_b ( t_a )]}
 = G_{kj}^p ( t' , t )\, , \\
G_{kk}^p ( t , t' ) & = & \Theta ( t - t' ) \frac{[ \dot{D}_b ( t ) 
\dot{D}_a ( t_b )+ \dot{D}_a ( t )] [ \dot{D}_a ( t' ) \dot{D}_b ( t_a ) - 
\dot{D}_b ( t' )]}{D_a ( t_b )[1 + 
\dot{D}_a( t_b ) \dot{D}_b ( t_a )]} \nonumber \\ 
& & + \Theta ( t' - t ) \frac{
[ \dot{D}_a ( t ) \dot{D}_b ( t_a ) - 
\dot{D}_b ( t )][ \dot{D}_b ( t' ) \dot{D}_a ( t_b )
+ \dot{D}_a ( t' )]}{D_a ( t_b )[1 + 
\dot{D}_a( t_b ) \dot{D}_b ( t_a )]} 
 = G_{kk}^p ( t' , t )
\, .
\end{eqnarray}
The relation to the correlation functions is similar to 
(\ref{59a})--(\ref{59c}):
\begin{eqnarray}
  \label{63a}
  \langle \,p_b\,|\,\tx(t)\,\tx(t')\,|\,p_a\,\rangle &=&\frac{i\hbar}{M}G_{jj}^p(t,t'),\\
  \label{63b}
  \langle \,p_b\,|\,\tx(t)\,\tp(t')\,|\,p_a\,\rangle &=&i\hbar G_{jk}^p(t,t')=i\hbar G_{kj}^p(t',t),\\
  \label{63c}
  \langle \,p_b\,|\,\tp(t)\,\tp(t')\,|\,p_a\,\rangle &=&i\hbar MG_{kk}^p(t,t').
\end{eqnarray}
The relation between the similar-looking actions (\ref{57b}) and (\ref{61}) 
becomes more transparent by reexpressing both
in terms of partial derivatives of the
classical solutions $x_{\rm cl} ( t ), \bar{x}_{\rm cl} ( t )$, 
$p_{\rm cl} ( t ), \bar{p}_{\rm cl} ( t )$
with respect to the end points $x_b, x_a$ and $p_b, p_a$, respectively.
In the configuration representation we obtain
\begin{eqnarray}
&&{\cal A} ( x_b , t_b ; x_a , t_a ) [ k ( t ) , j ( t )]
= \frac{1}{2} \left( x_b , x_a \right) \, \left( 
\begin{array}{rr}
{\displaystyle\frac{\partial p_b}{\partial x_b}} & \hspace*{0.2cm}
{\displaystyle\frac{\partial p_b}{\partial x_a}} \\[2mm]
- {\displaystyle\frac{\partial p_a}{\partial x_b}} &\hspace*{0.2cm}
- {\displaystyle\frac{\partial p_a}{\partial x_a}} 
\end{array}
\right) \,
\left( \begin{array}{c} x_b \\ x_a \end{array} \right) \nonumber \\
&& \hspace*{1cm} + \int\limits_{t_a}^{t_b} d t \, 
\left( x_b , x_a \right)
\, \left( 
\begin{array}{rr}
{\displaystyle\frac{\partial p_{\rm cl} ( t )}{\partial x_b}} & \hspace*{0.2cm}
{\displaystyle\frac{\partial x_{\rm cl} ( t )}{\partial x_b}} \\[2mm]
{\displaystyle\frac{\partial p_{\rm cl} ( t )}{\partial x_a}} &\hspace*{0.2cm}
{\displaystyle\frac{\partial x_{\rm cl} ( t )}{\partial x_a}} 
\end{array}
\right) \,\left( \begin{array}{c} k ( t )  \\ j ( t ) \end{array} 
\right) \nonumber\\
&& \hspace*{1cm}
- \frac{1}{2} \frac{\partial x_b}{\partial p_a} 
\int\limits_{t_a}^{t_b} d t \int\limits_{t_a}^{t_b} d t'
\left( k ( t )  , j ( t ) \right) \left[ \Theta ( t - t' )
\left( 
\begin{array}{rr}
{\displaystyle \frac{\partial p_{\rm cl} ( t )}{\partial x_a} 
\frac{\partial p_{\rm cl} ( t' )}{\partial x_b}} & \hspace*{0.2cm}
{\displaystyle\frac{\partial p_{\rm cl} ( t )}{\partial x_a} 
\frac{\partial x_{\rm cl} ( t' )}{\partial x_b}} \\[2mm]
{\displaystyle\frac{\partial x_{\rm cl} ( t )}{\partial x_a} 
\frac{\partial p_{\rm cl} ( t' )}{\partial x_b}} &\hspace*{0.2cm}
{\displaystyle\frac{\partial x_{\rm cl} ( t )}{\partial x_a} 
\frac{\partial x_{\rm cl} ( t' )}{\partial x_b}} 
\end{array}
\right) \right. \nonumber \\ 
&& \hspace*{1cm}
\left. + \Theta ( t' - t )
\left( 
\begin{array}{rr}
{\displaystyle\frac{\partial p_{\rm cl} ( t )}{\partial x_b} 
\frac{\partial p_{\rm cl} ( t' )}{\partial x_a}} & \hspace*{0.2cm}
{\displaystyle\frac{\partial p_{\rm cl} ( t )}{\partial x_b} 
\frac{\partial x_{\rm cl} ( t' )}{\partial x_a}} \\[2mm]
{\displaystyle\frac{\partial x_{\rm cl} ( t )}{\partial x_b} 
\frac{\partial p_{\rm cl} ( t' )}{\partial x_a}} &\hspace*{0.2cm}
{\displaystyle\frac{\partial x_{\rm cl} ( t )}{\partial x_b} 
\frac{\partial x_{\rm cl} ( t' )}{\partial x_a}} 
\end{array}
\right) \right]
 \,\left( \begin{array}{c} k ( t' )  \\ j ( t' ) \end{array} \right) 
 \, .
\end{eqnarray}
The momentum representation, on the other hand, has the analogues form with
$x$ and $p$ interchanged:
\begin{eqnarray}
&&{\cal A} ( p_b , t_b ; p_a , t_a ) [ k ( t ) , j ( t )]
= \frac{1}{2} \left( p_b , p_a \right) \, \left( 
\begin{array}{rr}
- {\displaystyle\frac{\partial x_b}{\partial p_b}} & \hspace*{0.2cm}
- {\displaystyle\frac{\partial x_b}{\partial p_a}} \\[2mm]
{\displaystyle\frac{\partial x_a}{\partial p_b}} &\hspace*{0.2cm}
{\displaystyle\frac{\partial x_a}{\partial p_a}} 
\end{array}
\right) \,
\left( \begin{array}{c} p_b \\ p_a \end{array} \right) \nonumber \\
&& \hspace*{1cm}
+ \int\limits_{t_a}^{t_b} d t \, 
\left( p_b , p_a \right)
\, \left( 
\begin{array}{rr}
{\displaystyle\frac{\partial \bar{p}_{\rm cl} ( t )}{\partial p_b}} &
\hspace*{0.2cm}
{\displaystyle\frac{\partial \bar{x}_{\rm cl} ( t )}{\partial p_b}} \\[2mm]
{\displaystyle\frac{\partial \bar{p}_{\rm cl} ( t )}{\partial p_a}} &
\hspace*{0.2cm}
{\displaystyle\frac{\partial \bar{x}_{\rm cl} ( t )}{\partial p_a}} 
\end{array}
\right) \,\left( \begin{array}{c} k ( t )  \\ j ( t ) \end{array} 
\right) \nonumber\\
&& \hspace*{1cm}
- \frac{1}{2} \frac{\partial x_b}{\partial p_a} 
\int\limits_{t_a}^{t_b} d t \int\limits_{t_a}^{t_b} d t'
\left( k ( t )  , j ( t ) \right) \left[ \Theta ( t - t' )
\left( 
\begin{array}{rr}
{\displaystyle \frac{\partial \bar{p}_{\rm cl} ( t )}{\partial p_a} 
\frac{\partial \bar{p}_{\rm cl} ( t' )}{\partial p_b}} & \hspace*{0.2cm}
{\displaystyle\frac{\partial \bar{p}_{\rm cl} ( t )}{\partial p_a} 
\frac{\partial \bar{x}_{\rm cl} ( t' )}{\partial p_b}} \\[2mm]
{\displaystyle\frac{\partial \bar{x}_{\rm cl} ( t )}{\partial p_a} 
\frac{\partial \bar{p}_{\rm cl} ( t' )}{\partial p_b}} &\hspace*{0.2cm}
{\displaystyle\frac{\partial \bar{x}_{\rm cl} ( t )}{\partial p_a} 
\frac{\partial \bar{x}_{\rm cl} ( t' )}{\partial p_b}} 
\end{array}
\right) \right. \nonumber \\ 
&& \hspace*{1cm}
\left. + \Theta ( t' - t )
\left( 
\begin{array}{rr}
{\displaystyle\frac{\partial \bar{p}_{\rm cl} ( t )}{\partial p_b} 
\frac{\partial \bar{p}_{\rm cl} ( t' )}{\partial p_a}} & \hspace*{0.2cm}
{\displaystyle\frac{\partial \bar{p}_{\rm cl} ( t )}{\partial p_b} 
\frac{\partial \bar{x}_{\rm cl} ( t' )}{\partial p_a}} \\[2mm]
{\displaystyle\frac{\partial \bar{x}_{\rm cl} ( t )}{\partial p_b} 
\frac{\partial \bar{p}_{\rm cl} ( t' )}{\partial p_a}} &\hspace*{0.2cm}
{\displaystyle\frac{\partial \bar{x}_{\rm cl} ( t )}{\partial p_b} 
\frac{\partial \bar{x}_{\rm cl} ( t' )}{\partial p_a}} 
\end{array}
\right) \right]
 \,\left( \begin{array}{c} k ( t' )  \\ j ( t' ) \end{array} \right) 
 \, .
\end{eqnarray}
These expressions for the generating functionals (\ref{57}) and (\ref{60b}) 
exhibit clearly the symmetry properties (\ref{10}) and (\ref{11}).
\section{Smearing Formula for Harmonic Fluctuations}
\label{harmfluct}
As a first application of the generating functional (\ref{57}) 
we derive a general rule for 
calculating correlation functions of polynomial or nonpolynomial functions of $x ( t )$ and $p ( t )$.
The result will be expressed in the form of a {\it smearing formula}. 
This formula will represent an essential tool for 
calculating perturbation expansions with nonpolynomial interactions.
Such expansions serve in variational perturbation theory to obtain convergent approximations for quantum-statistical partition functions \cite{Werner} or density matrices \cite{bkp}.

Consider the correlation functions of 
a product of local functions for vanishing currents
\begin{eqnarray}
  \label{sm01}
&&\langle\,F_1(x(t_1))\,F_2(x(t_2))\ldots F_N(x(t_N))\,F_{N+1}
(p(t_{N+1}))\,F_{N+2}(p(t_{N+2}))\ldots F_{N+M}(p(t_{N+M}))\,
\rangle_\Omega^{x_b,x_a}\nonumber\\
& & = \frac{1}{(\,x_b\,t_b\,|\,x_a\,t_a\,)}\,\int\limits_{x_a,t_a}^{x_b,t_b}
\frac{{\cal D}x\,{\cal D}p}{2\pi\hbar}\,\prod\limits_{n=1}^N\left[F_n(x(t_n)) 
\right]\prod\limits_{m=1}^M\left[F_{N+m}(p(t_{N+m})) \right]\,
\exp\left\{\frac{i}{\hbar}{\cal A}[p,x;0,0]\right\},
\end{eqnarray}
where the harmonic time evolution amplitude with zero external currents $(x_b\,t_b\,|\,x_a\,t_a)[0,0]$ is written as $(x_b\,t_b\,|\,x_a\,t_a)$.
By Fourier transforming
the functions $F_n(x(t_n))$ and $F_{N+m}(p(t_{N+m}))$ according to
\begin{equation}
  \label{sm03a}
  F_n(x(t_n))=\int\limits_{-\infty}^{+\infty}dx_n\,F_n(x_n)
\delta(x_n-x(t_n))=\int\limits_{-\infty}^{+\infty}dx_n\,F(x_n)
\int\limits_{-\infty}^{+\infty}\frac{d\xi_n}{2\pi}\,\exp
\left\{i\xi_n(x_n-x(\tau_n)) \right\}
\end{equation}
and
\begin{eqnarray}
  \label{sm03b}
 && F_{N+m}(p(t_{N+m}))=\nonumber\\
&&\hspace*{0.4cm}\int\limits_{-\infty}^{+\infty}
\frac{dp_m}{2\pi\hbar}F_{N+m}(p_m)\delta(p_m-p(t_{N+m}))
=\int\limits_{-\infty}^{+\infty}\frac{dp_m}{2\pi\hbar}F_{N+m}(p_m)
\int\limits_{-\infty}^{+\infty}d\kappa_m\,e^{-i\kappa_m(p_m-p(t_{N+m}))/\hbar },
\end{eqnarray}
the correlation functions (\ref{sm01}) may be reexpressed as
\begin{eqnarray}
  \label{sm04}
\lefteqn{\langle\,F_1(x(t_1))\ldots F_{N+M}(p(t_{N+M}))\,
\rangle_\Omega^{x_b,x_a} = \frac{1}{(\,x_b\,t_b\,|\,x_a\,t_a\,)}
\prod\limits_{n=1}^N\left[\int\limits_{-\infty}^{+\infty}dx_n\,F_n(x_n) 
\int\limits_{-\infty}^{+\infty}\frac{d\xi_n}{2\pi}\,e^{i\xi_nx_n}\right]}
\hspace{80pt}\nonumber\\
& &\times\prod\limits_{m=1}^M\left[\int\limits_{-\infty}^{+\infty}
\frac{dp_m}{2\pi\hbar}\,F_{N+m}(p_m)\int\limits_{-\infty}^{+\infty}
d\kappa_m e^{-i\kappa_mp_m/\hbar} \right]\,(\,x_b\,t_b\,|\,x_a\,t_a\,)[k,j],
\end{eqnarray}
where the generating functional is given by (\ref{57}). The 
currents $j(t)$ and $k(t)$ are specialized to
\begin{equation}
  \label{sm05}
  j(t) = -\hbar\sum\limits_{n=1}^N\,\xi_n\delta(t-t_n),\qquad 
k(t)=\sum\limits_{m=1}^M\,\kappa_m\delta(t-t_{N+m}).
\end{equation}
Inserting these equations into the 
action (\ref{57b}) and the Green functions (\ref{37}), 
(\ref{58}) and (\ref{59}), we find the Fourier decomposition of the
the generating functional (\ref{57}), so that the correlation functions
(\ref{sm04}) become
\begin{eqnarray}
  \label{sm06}
  \lefteqn{\langle\,F_1(x(t_1))\ldots F_{N+M}(p(t_{N+M}))\,
\rangle_\Omega^{x_b,x_a} =\prod\limits_{n=1}^N\left[
\int\limits_{-\infty}^{+\infty}dx_n\,F_n(x_n) \int\limits_{-\infty}^{+\infty}
\frac{d\xi_n}{2\pi}\,e^{i\xi_n(x_n-x_{\rm cl}(t_n))}\right]}\hspace{20pt}
\nonumber\\ 
&\times&\prod\limits_{m=1}^M\left[\int\limits_{-\infty}^{+\infty}
\frac{dp_m}{2\pi\hbar}\,F_{N+m}(p_m)\int\limits_{-\infty}^{+\infty}
d\kappa_m e^{-i\kappa_m(p_m-p_{\rm cl}(t_{N+m}))/\hbar} \right]\nonumber\\
&\times&\exp\left\{-\frac{i\hbar}{2M}\sum\limits_{n,n'=1}^N\,
\xi_nG_{jj}^{n,n'}\xi_{n'} +i\sum\limits_{n=1}^N\sum_{m=1}^M\,\xi_n
G_{jk}^{nm}\kappa_m-\frac{iM}{2\hbar}\sum\limits_{m,m'=1}^M\,
\kappa_mG_{kk}^{mm'}\kappa_{m'}\right\},
\end{eqnarray}
where we used the abbreviations
\begin{equation}
  \label{sm07}
  G_{jj}^{nn'}=G_{jj}^x(t_n,t_{n'}),\quad G_{jk}^{nm}=G_{jk}^x(t_n,t_{N+m})
,\quad G_{kk}^{mm'}=G_{kk}^x(t_{N+m},t_{N+m'}).
\end{equation}
To proceed, 
it is more convenient to write expression (\ref{sm06}) as a convolution 
integral
\begin{eqnarray}
  \label{sm08}
  \langle\,F_1(x(t_1))\ldots F_{N+M}(p(t_{N+M}))\,\rangle_\Omega^{x_b,x_a} 
&=&\prod\limits_{n=1}^N\left[\int_{-\infty}^{+\infty}dx_n\,F_n(x_n)\right]
\prod\limits_{m=1}^M\left[\int_{-\infty}^{+\infty}\frac{dp_m}{2\pi\hbar}\,
F_{N+m}(p_m)\right]\nonumber\\
& &\times \left(\frac{M\Omega}{\hbar}\right)^{(N-M)/2}\,P(x_1,\ldots,x_N,
p_{1},\ldots,p_{M})
\end{eqnarray}
involving the Gaussian distribution
\begin{equation}
  \label{sm09}
  P(x_1,\ldots,p_{M}) \equiv \frac{1}{(2\pi)^N}\int d^{N+M}v\,\exp
\left\{i{\bf w}^T{\bf v}-\frac{i}{2}{\bf v}^T\,G\,{\bf v}\right\}.
\end{equation}
The dimensionless vectors ${\bf v}$ and ${\bf w}$ have $N+M$ components and
are defined as
\begin{equation}
  \label{sm10a}
  {\bf v}^T=\left(\sqrt{\frac{\hbar}{M\Omega}}\xi_1,\ldots,
\sqrt{\frac{\hbar}{M\Omega}}\xi_N,\sqrt{\frac{M\Omega}{\hbar}}
\kappa_1,\ldots,\sqrt{\frac{M\Omega}{\hbar}}\kappa_M\right)
\end{equation}
and
\begin{eqnarray}
  \label{sm10b}
  {\bf w}^T&=&\Bigg(\sqrt{\frac{M\Omega}{\hbar}}\left[ x_1-x_{\rm cl}(t_1)
\right],
\ldots,\sqrt{\frac{M\Omega}{\hbar}}\left[x_N-x_{\rm cl}(t_N)\right],\nonumber\\
& &-\frac{1}{\sqrt{\hbar M\Omega}}\left[p_1-p_{\rm cl}(t_{N+1})\right],\ldots,-
\frac{1}{\sqrt{\hbar M\Omega}}(p_M-p_{\rm cl}(t_{N+M})) \Bigg).
\end{eqnarray}
The $(N+M)\times (N+M)$-matrix of Green functions
\begin{equation}
  \label{sm11a}
  G=\left(\begin{array}{cc}A & B\\ B^T & C \end{array}\right)
\end{equation}
can be decomposed into block matrices $A$, $B$, and $C$. The $N \times 
N$-matrix $A$ and the $M \times M$-matrix $C$ are defined by
\begin{equation}
  \label{sm11b}
  A=\Omega\left(\begin{array}{cccc}
G_{jj}^{11} & G_{jj}^{12} & \cdots & G_{jj}^{1N}\\ 
G_{jj}^{12} & G_{jj}^{11} & \cdots & G_{jj}^{2N}\\
\vdots      & \vdots      & \ddots & \vdots\\
G_{jj}^{1N} & G_{jj}^{2N} & \cdots & G_{jj}^{11}
\end{array} \right), \hspace*{0.5cm}
C=\frac{1}{\Omega}\left(\begin{array}{cccc}
G_{kk}^{11} & G_{kk}^{12} & \cdots & G_{kk}^{1M}\\ 
G_{kk}^{12} & G_{kk}^{11} & \cdots & G_{kk}^{2M}\\
\vdots      & \vdots      & \ddots & \vdots\\
G_{kk}^{1M} & G_{kk}^{2M} & \cdots & G_{kk}^{11}
\end{array} \right)
\end{equation}
and yield quadratic forms of the position and momentum variables, respectively. The $N \times M$-matrix
\begin{equation}
  \label{sm11c}
  B=\left(\begin{array}{cccc}
- G_{jk}^{11} & - G_{jk}^{12} & \cdots & - G_{jk}^{1M}\\ 
- G_{jk}^{21} & - G_{jk}^{11} & \cdots & - G_{jk}^{2M}\\
\vdots        & \vdots        & \vdots & \vdots\\
- G_{jk}^{N1} & - G_{jk}^{N2} & \cdots & - G_{jk}^{NM}
\end{array} \right)
\end{equation}
gives rise to quadratic terms which are linear in both position and momentum
variables. The multidimensional integral in (\ref{sm09}) is of 
Fresnel type and can 
easily be done, yielding and explicit expression for 
the Gaussian distribution (\ref{sm08})
\begin{equation}
  \label{sm12}
  P(x_1,\ldots\,x_N,p_1,\ldots,p_M)=\frac{1}{\sqrt{i^{N+M}(2
\pi)^{N-M}{\rm det}\,G}}\exp\left\{\frac{i}{2}{\bf w}^T\,G^{-1}\,{\bf w} 
\right\} \, ,
\end{equation}
where $G^{-1}$ represents the matrix inverse of (\ref{sm11a}) whose block form
is
\begin{equation}
  \label{sm13}
  G^{-1}=\left(\begin{array}{cc}X^{-1} & -X^{-1}BC^{-1}\\-C^{-1}B^TX^{-1} 
& C^{-1}+C^{-1}B^TX^{-1}BC^{-1} \end{array}\right)
\end{equation}
with the abbreviation
\begin{equation}
  \label{sm14a}
  X=A-BC^{-1}B^T.
\end{equation}
Since the matrix $G$ may be decomposed as
\begin{eqnarray}
G = \left( \begin{array}{@{}cc} 1 & B \\ 0 & C \end{array} \right) 
\, \left( \begin{array}{@{}cc} X & 0 \\ C^{-1} B^T & 1 \end{array} 
\right) 
\end{eqnarray}
when the matrix $C$ is regular, the determinant of $G$ factorizes as follows
\begin{equation}
  \label{sm14b}
  \det G=\det C\,\det X.
\end{equation}
For singular matrix $C$ but $A$ regular, one may make use of another decomposition,
\begin{eqnarray}
\label{sm15}
G = \left( \begin{array}{@{}cc} 1 & 0 \\ B^TA^{-1} & X' \end{array} \right) 
\, \left( \begin{array}{@{}cc} A & B \\ 0 & 1 \end{array} 
\right) \, ,
\end{eqnarray}
with $X'=C-B^TA^{-1}B$. Then the determinant of $G$ is given by
\begin{equation}
  \label{sm16}  \det G=\det X'\,\det A.
\end{equation}
With the Gaussian distribution (\ref{sm12}), our result (\ref{sm08})
constitutes a {\it smearing formula} which describes the effect of
harmonic fluctuations upon arbitrary products of functions of space and
momentum variables at different times. 
\section{Generalized Wick Rules and Feynman Diagrams}
In applications, there often occur correlation functions for mixtures
of nonpolynomial functions $F(\tx(t))$ or $F(\tp(t))$ and powers 
according to
\begin{eqnarray}
  \mean{F(\tx(t_1))\,\tx^n(t_2)},\,\mean{F(\tx(t_1))\,\tp^n(t_2)},\, 
\nonumber \\
\mean{F(\tp(t_1))\,\tx^n(t_2)},\,\mean{F(\tp(t_1))\,\tp^n(t_2)} \, .
  \label{der01a}
\end{eqnarray}
In order to evaluate such correlation functions,
we derive in this section generalized Wick rules and Feynman diagrams
on the basis of the smearing formula (\ref{sm08}).
\subsection{Ordinary Wick Rules}
It is well known that if one has to calculate expectation values of 
polynomials with even power, Wick's rule can be written as the sum over 
all possible permutations of products of two-point functions. We shortly 
recall to this expansion by considering the case of a position-dependent 
$n$-point correlation function, $n$ even, defined as
\begin{equation}
  \label{der00a}
  G^{(n)}(t_1,\ldots,t_n)=\mean{\tx(t_1)\cdots \tx(t_n)}.
\end{equation}
Note that it will be sufficient to study only the correlation functions
involving the deviations from the classical path, respectively.
This expectation value can be decomposed with the help of Wick's expansion
\begin{equation}
  \label{der00b}
  G^{(n)}(t_1,\ldots,t_n)=\sum\limits_{\rm pairs}\,G^{(2)}(t_{p(1)},t_{p(2)})\cdots G^{(2)}(t_{p(n-1)},t_{p(n)}),
\end{equation}
where $p$ denotes the operation of pairwise index permutation. Thereby, the Green function $G^{(2)}(t_1,t_2)$ is already given by (\ref{59a}).
Note that Eq.~(\ref{der00b}) may be considered as a consequence of a simple derivative rule
\begin{equation}
  \label{der00d}
  \mean{F(\tx(t_1))\,\tx(t_2)}=\mean{\tx(t_1)\,\tx(t_2)}\mean{F'(\tx(t_1))}
\end{equation}
with $F'(\tx)=\partial F(\tx)/\partial x$. By applying this recursively, one 
eventually obtains (\ref{der00b}). And conversely, the derivative rule 
(\ref{der00d}) can be proved for {\it polynomial} functions $F(\tx(t))$, 
following directly from Wick's theorem (\ref{der00b}). 

The two-point Green functions $G^{(2)}(t_1,t_2)$, occuring in (\ref{der00b}), can be considered as a Wick contraction which we introduce as follows:
\setlength{\unitlength}{1cm}
\begin{eqnarray}
\label{w01a}
\parbox{10.5cm}
{\begin{picture}(10.5,0.9)
\put(0,0.3){$\displaystyle{\tx(t_1)\,\tx(t_2)=\mean{\tx(t_1)\,\tx(t_2)}={\frac{i\hbar}{M}}G_{jj}(t_1,t_2)},$}
\put(0.4,0){\line(1,0){1.0}}
\put(0.4,0){\line(0,1){0.15}}
\put(1.4,0){\line(0,1){0.15}}
\end{picture}}\\
\label{w01b}
\parbox{10.5cm}{
\begin{picture}(10.5,0.9)
\put(0,0.3){$\displaystyle{\tx(t_1)\,\tp(t_2)=\mean{\tx(t_1)\,\tp(t_2)}=i\hbar G_{jk}(t_1,t_2)},$}
\put(0.4,0){\line(1,0){1.0}}
\put(0.4,0){\line(0,1){0.15}}
\put(1.4,0){\line(0,1){0.15}}
\end{picture}}\\
\label{w01c}
\parbox{10.5cm}{
\begin{picture}(10.5,0.9)
\put(0,0.3){$\displaystyle{\tp(t_1)\,\tx(t_2)=\mean{\tp(t_1)\,\tx(t_2)}=i\hbar G_{kj}(t_1,t_2)=i\hbar G_{jk}(t_2,t_1)},$}
\put(0.4,0){\line(1,0){1.0}}
\put(0.4,0){\line(0,1){0.15}}
\put(1.4,0){\line(0,1){0.15}}
\end{picture}}\\
\label{w01d}
\parbox{10.5cm}{
\begin{picture}(10.5,0.9)
\put(0,0.3){$\displaystyle{\tp(t_1)\,\tp(t_2)=\mean{\tp(t_1)\,\tp(t_2)}=i\hbar M G_{kk}(t_1,t_2)}.$}
\put(0.4,0){\line(1,0){1.0}}
\put(0.4,0){\line(0,1){0.15}}
\put(1.4,0){\line(0,1){0.15}}
\end{picture}}
\end{eqnarray}
Decomposing polynomial correlations of $\tx(t)$ and $\tp(t)$ with the help of these contractions corresponding to Eq.~(\ref{der00b}) or successively applying the derivative rule (\ref{der00d}) leads to following results
\begin{eqnarray}
  \label{w02a}
&&\hspace{-15pt}\mean{\tx^n(t_1)\,\tx^m(t_2)}=\nonumber\\
&&\sum\limits_{l=\alpha,\alpha+2,\alpha+4,\ldots}^{\min(n,m)}\,c_l\,\left[\frac{i\hbar}{M}G_{jj}(t_1,t_1)\right]^{(n-l)/2}\left[\frac{i\hbar}{M} G_{jj}(t_1,t_2)\right]^l\left[\frac{i\hbar}{M}G_{jj}(t_2,t_2)\right]^{(m-l)/2},\\
  \label{w02b}
&&\hspace{-15pt}\mean{\tx^n(t_1)\,\tp^m(t_2)}=\nonumber\\
&&\sum\limits_{l=\alpha,\alpha+2,\alpha+4,\ldots}^{\min(n,m)}\,c_l\,\left[\frac{i\hbar}{M}G_{jj}(t_1,t_1)\right]^{(n-l)/2}[i\hbar G_{jk}(t_1,t_2)]^l\left[i\hbar MG_{kk}(t_2,t_2)\right]^{(m-l)/2},\\
  \label{w02c}
&&\hspace{-15pt}\mean{\tp^n(t_1)\,\tx^m(t_2)}=\nonumber\\
&&\sum\limits_{l=\alpha,\alpha+2,\alpha+4,\ldots}^{\min(n,m)}\,c_l\,\left[i\hbar MG_{kk}(t_1,t_1)\right]^{(n-l)/2}[i\hbar G_{jk}(t_2,t_1)]^l\left[\frac{i\hbar}{M}G_{jj}(t_2,t_2)\right]^{(m-l)/2},\\
  \label{w02d}
&&\hspace{-15pt}\mean{\tp^n(t_1)\,\tp^m(t_2)}=\nonumber\\
&&\sum\limits_{l=\alpha,\alpha+2,\alpha+4,\ldots}^{\min(n,m)}\,c_l\,\left[i\hbar MG_{kk}(t_1,t_1)\right]^{(n-l)/2}\left[i\hbar M G_{kk}(t_1,t_2)\right]^l\left[i\hbar MG_{kk}(t_2,t_2)\right]^{(m-l)/2}
\end{eqnarray}
with the multiplicity factor
\begin{equation}
  \label{w03}
  c_l=\frac{(n-l-1)!!\,(m-l-1)!!\,n!\,m!}{l!\,(n-l)!\,(m-l)!}.
\end{equation}
Note, that $(-k)!!\equiv 1$ for any positive integer $k$. For nonvanishing correlation, the sum $n+m$ must be even so that the regulation parameter $\alpha$ is defined as follows:
\begin{equation}
  \label{w04}
  \alpha=\left\{\begin{array}{ll}0,& \quad n,m\;\mbox{even},\\ 1,& \quad n,m\,\mbox{odd}.  \end{array} \right.
\end{equation}
The contractions defined in (\ref{w01a})-(\ref{w01d}) can be used to treat
The desired derivative 
rules read
\begin{eqnarray}
  \label{der06a}
  &&\hspace{-5pt} \mean{F(\tx(t_1))\,\tx^n(t_2)}=\nonumber\\
&&\sum\limits_{l=\alpha,\alpha+2,\alpha+4,\ldots}^n\,
\frac{n!}{(n-l)!!\,l!}\,\left[\frac{i\hbar}{M}G_{jj}
(t_2,t_2)\right]^{(n-l)/2}\,\left[\frac{i\hbar}{M} 
G_{jj}(t_1,t_2)\right]^l\,\mean{F^{(l)}(\tx(t_1))},\\
\label{der06b}
 &&\hspace{-10pt} \mean{F(\tx(t_1))\,\tp^n(t_2)}=\nonumber\\
&&\sum\limits_{l=\alpha,\alpha+2,\alpha+4,\ldots}^n\,
\frac{n!}{(n-l)!!\,l!}\,[i\hbar MG_{kk}(t_2,t_2)]^{(n-l)/2}\,\
[i\hbar G_{jk}(t_1,t_2)]^l\,\mean{F^{(l)}(\tx(t_1))},\\
\label{der06c}
&&\hspace{-5pt} \mean{F(\tp(t_1))\,\tp^n(t_2)}=\nonumber\\
&&\sum\limits_{l=\alpha,\alpha+2,\alpha+4,\ldots}^n\,
\frac{n!}{(n-l)!!\,l!}\,[i\hbar MG_{kk}(t_2,t_2)]^{(n-l)/2}\,
[i\hbar MG_{kk}(t_1,t_2)]^l\,\mean{F^{(l)}(\tp(t_1))},\\
\label{der06d}
&&\hspace{-5pt} \mean{F(\tp(t_1))\,\tx^n(t_2)}=\nonumber\\
&&\sum\limits_{l=\alpha,\alpha+2,\alpha+4,\ldots}^n\,
\frac{n!}{(n-l)!!\,l!}\,\left[\frac{i\hbar}{M}G_{jj}(t_2,t_2)
\right]^{(n-l)/2}\,[i\hbar G_{jk}(t_2,t_1)]^l\,\mean{F^{(l)}(\tp(t_1))}.
\end{eqnarray}
The parameter $\alpha$ distinguishes between even and odd power $n$:
\begin{equation}
  \label{der05b}
  \alpha=\left\{\begin{array}{ll}0,& \quad n\;\mbox{even},\\ 1,& \quad n\;\mbox{odd,}  \end{array} \right.
\end{equation}
since even (odd) powers of $n$ lead to even (odd) derivatives of the function $F(\tx(t_1))$.
The $l$th derivative $F^{(l)}(\tx(t_1))$ is formed with respect to $x(t_1)$, and $F^{(l)}(\tp(t_1))$ is the $l$th derivative with respect to $p(t_1)$. Note, that in the last line the Green function $G_{jk}$ appears with exchanged time arguments, which in this case happens to be inessential due to the symmetry $G_{jk}(t_2,t_1)=G_{kj}(t_1,t_2)$. 

\subsection{Generalized Wick Rule}
\label{wick}
According to their derivation, the contractions (\ref{der06a})-(\ref{der06d})
are only applicable to functions $F ( \tx ( t ) )$ and $F ( \tp ( t ) )$
which can be Taylor-expanded. In the following, we will show with the help
of the smearing formula (\ref{sm08}) that these derivative rules remain
valid for functions $F ( \tx ( t ) )$ and $F ( \tp ( t ) )$ with 
Laurent expansions.
Expectations of this type appear in variational perturbation theory 
(see for position-position coupling Ref.~\cite{bkp}).
Since the proceeding is similar in all the cases 
(\ref{der06a})-(\ref{der06d}), 
we shall only discuss the expectation value 
\begin{equation}
  \label{der01b}
  \mean{F(\tx(t_1))\,\tp^n(t_2)}
\end{equation}
in detail. For this we consider the generating functional of all such 
expectation values following from (\ref{sm08})
\begin{eqnarray}
  \label{der03}
  &&\hspace{-20pt}\mean{F(\tx(t_1))\,e^{j\tp^n(t_2)}}=\frac{1}{\sqrt{-\det G}}\int\limits_{-\infty}^{+\infty}dx\,F(x)\int\limits_{-\infty}^{+\infty}\frac{dp}{2\pi\hbar}e^{jp}\nonumber\\
&&\times\exp\left\{\frac{i}{2\det G}\left[\frac{M}{\hbar}G_{kk}(t_2,t_2)\,x^2-2\frac{1}{\hbar}G_{jk}(t_1,t_2)\,xp+\frac{1}{\hbar M}G_{jj}(t_1,t_1)\,p^2 \right] \right\}.
\end{eqnarray}
The $p$-integration can easily be done, leading to
\begin{eqnarray}
  \label{der04}
  \mean{F(\tx(t_1))\,e^{j\tp^n(t_2)}}&=&e^{i\hbar MG_{kk}(t_2,t_2)j^2/2}\nonumber\\
&&\times\int\limits_{-\infty}^{+\infty}\frac{dx}{\sqrt{2\pi i\hbar G_{jj}(t_1,t_1)/M}}\,F(x+i\hbar G_{jk}(t_1,t_2)\,j)\,e^{iMx^2/2\hbar G_{jj}(t_1,t_1)}\nonumber\\
&=&e^{i\hbar MG_{kk}(t_2,t_2)j^2/2}\,\sum_{l=0}^\infty\,\frac{1}{l!}\,[i\hbar G_{jk}(t_1,t_2)\,j]^l\,\mean{F^{(l)}(\tx(t_1))}.
\end{eqnarray}
The correlation of two functions at different times has been reduced to a 
single-time expectation value of the $l$th derivative of the function 
$F(\tx(t_1))$ with respect to $x(t_1)$, denoted by $F^{(l)}(\tx(t_1))$, 
with Green functions describing the dependence on the second time. 
Expanding both sides in powers of $j$, we reobtain (\ref{der06b}). 

Now we demonstrate that the derivative rules (\ref{der06a})-(\ref{der06d})
for Laurent-expandable functions $F ( \tx ( t ) )$ and $F ( \tp ( t ) )$ 
also follow from generalized Wick rules. Without restriction of universality,
we only consider the expectation value

\begin{equation}
\label{w05}
\mean{F(\tx(t_1))\,\tx^n(t_2)}.
\end{equation}
The proceeding to reduce the power of the polynomial at the expense of the function $F(\tx(t_1))$ is as follows:\\
{\bf 1a.{}} If possible ($n\ge 2$), contract $\tx(t_2)\,\tx(t_2)$ with multiplicity $(n-1)$, giving
\begin{equation}
  \label{w06a}
\parbox{10.5cm}
{\begin{picture}(10.5,0.6)
\put(0,0.3){$(n-1)\,\tx(t_2)\,\tx(t_2)\,\mean{F(\tx(t_1))\,\tx^{n-2}(t_2)},$}
\put(1.8,0){\line(1,0){0.9}}
\put(1.8,0){\line(0,1){0.15}}
\put(2.7,0){\line(0,1){0.15}}
\end{picture}}
\end{equation}
else jump to 1b. directly.\\
{\bf 1b.{}} Contract $F(\tx(t_1))\,\tx(t_2)$ and let the remaining polynomial invariant. We define this contraction by the symbol
\begin{equation}
\label{w06b}
\parbox{10.5cm}
{\begin{picture}(10.5,0.6)
\put(0,0.3){$\displaystyle{F(\tx(t_1))\,\tx(t_2)\,\tx^{n-1}(t_2)=\tx(t_1)\,\tx(t_2)\,\mean{F'(\tx(t_1))\,\tx^{n-1}(t_2)}}.$}
\put(0.8,0){\line(1,0){1.2}}
\put(0.8,0){\line(0,1){0.15}}
\put(2.0,0){\line(0,1){0.15}}
\multiput(2.0,0)(0.3,0){5}{\line(1,0){0.15}}
\put(3.35,0){\line(0,1){0.15}}
\put(5.0,0){\line(1,0){0.9}}
\put(5.0,0){\line(0,1){0.15}}
\put(5.9,0){\line(0,1){0.15}}
\end{picture}}
\end{equation}
{\bf 1c.{}} Add the terms 1a. and 1b.\\
{\bf 2.{}} Repeat steps 1a.-1c. until only expectation values of $F(\tx)$ or expectations of its derivatives remain.

Summarizing, we can express the first power reduction by the generalized Wick rule ($n\ge 2$)
\begin{eqnarray}
  \label{w06c}
  \parbox{10.5cm}
{\begin{picture}(10.5,0.6)
  \put(0,0.3){$\mean{F(\tx(t_1))\,\tx^n(t_2)}=(n-1)\,\tx(t_2)\,\tx(t_2)\,\mean{F(\tx(t_1))\,\tx^{n-2}(t_2)}$}
\put(6.0,0){\line(1,0){0.9}}
\put(6.0,0){\line(0,1){0.15}}
\put(6.9,0){\line(0,1){0.15}}
\end{picture}}\nonumber\\
  \parbox{10.5cm}
{\begin{picture}(10.5,0.6)
  \put(4.4,0.3){$+\,F(\tx(t_1))\,\tx(t_2)\,\tx^{n-1}(t_2)$}
\put(5.6,0){\line(1,0){1.2}}
\put(5.6,0){\line(0,1){0.15}}
\put(6.8,0){\line(0,1){0.15}}
\multiput(6.8,0)(0.3,0){5}{\line(1,0){0.15}}
\put(8.15,0){\line(0,1){0.15}}
\end{picture}}
\end{eqnarray}
with the contraction rules defined in (\ref{w01a}) and (\ref{w06b}). For $n=1$, we obtain
\begin{equation}
  \label{w06d}
  \parbox{10.5cm}
{\begin{picture}(10.5,0.6)
  \put(0,0.3){$\mean{F(\tx(t_1))\,\tx(t_2)}=\tx(t_1)\,\tx(t_2)\,\mean{F'(x(t_1))}$,}
\put(4.7,0){\line(1,0){0.9}}
\put(4.7,0){\line(0,1){0.15}}
\put(5.6,0){\line(0,1){0.15}}
\end{picture}}
\end{equation}
which is valid for {\it any} function $F(\tx(t))$ generalizing the rule 
(\ref{der00d}) that was proved for polynomial functions only. Recursively 
applying this power reduction, we finally end up with the derivative rule
(\ref{der06a}). Note that the generalization of Wick's rule for mixed 
position-momentum or pure momentum couplings is done along similar lines, 
leading to the derivative rules (\ref{der06b})-(\ref{der06d}).
\subsection{New Feynman-like Rules for Nonpolynomial Interactions}
\label{feynman}
Higher-order perturbation expressions become usually complicated. For simple polynomial interactions, Feynman diagrams are a useful tool to classify perturbative contributions with the help of graphical rules. Here, we are going to set up analogous diagrammatic rules for perturbation expansions for nonpolynomial interactions $V(x(t),p(t))$, whose contributions may be expressed as expectations values 
\begin{equation}
  \label{f00}
  \int\limits_{t_a}^{t_b}dt_n\cdots \int\limits_{t_a}^{t_b}dt_1\,\mean{V(x(t_n),p(t_n))\cdots V(x(t_1),p(t_1))}
\end{equation}
From (\ref{w01a})-(\ref{w01d}) follows that we have four basic propagators whose graphical representation may be defined as (setting $\hbar=M=1$ from now on)
\setlength{\unitlength}{1mm}
\begin{fmffile}{prop}
\begin{eqnarray}
\parbox{30mm}{\centerline{
\begin{fmfgraph*}(10,8)
\setval
\fmfleft{v1}
\fmfright{v2}
\fmf{plain}{v1,v2}\fmflabel{$t_1$}{v1}
\fmflabel{$t_2$}{v2}
\end{fmfgraph*}
}} &\equiv& \mean{\tx(t_1)\,\tx(t_2)}=iG_{jj}(t_1,t_2),\nonumber\\
\parbox{30mm}{\centerline{
\begin{fmfgraph*}(10,8)
\setval
\fmfleft{v1}
\fmfright{v2}
\fmf{wiggly}{v1,v2}
\fmflabel{$t_1$}{v1}
\fmflabel{$t_2$}{v2}
\end{fmfgraph*}
}} &\equiv& \mean{\tp(t_1)\,\tp(t_2)}=iG_{kk}(t_1,t_2), \nonumber\\
\parbox{30mm}{\centerline{
\begin{fmfgraph*}(10,8)
\setval
\fmfleft{v1}
\fmfright{v2}
\fmf{dashes_arrow}{v2,v1}
\fmflabel{$t_1$}{v1}
\fmflabel{$t_2$}{v2}
\end{fmfgraph*}
}} &\equiv& \mean{\tx(t_1)\,\tp(t_2)}=iG_{jk}(t_1,t_2), \nonumber\\
\parbox{30mm}{\centerline{
\begin{fmfgraph*}(10,8)
\setval
\fmfleft{v1}
\fmfright{v2}
\fmf{dashes_arrow}{v1,v2}
\fmflabel{$t_1$}{v1}
\fmflabel{$t_2$}{v2}
\end{fmfgraph*}
}} &\equiv& \mean{\tp(t_1)\,\tx(t_2)}=iG_{kj}(t_1,t_2)=iG_{jk}(t_2,t_1).\nonumber
\end{eqnarray}
A vertex is represented as usual by a small dot. The time variable is integrated over at a vertex in a perturbation expansion,
\begin{eqnarray}
\parbox{6mm}{\centerline{
\begin{fmfgraph}(5,8)
\setval
\fmfleft{v1}
\fmfdot{v1}
\end{fmfgraph}
}} \equiv \int_{t_a}^{t_b}dt\nonumber.
\end{eqnarray}
We now introduce the diagrammatic representations of the expectation value of arbitrary functions $F(\tx(t))$ or $F(\tp(t))$ and their derivatives as
\begin{center}
\begin{tabular}{lcl@{\hspace{2cm}}lcl}
\parbox{10mm}{\centerline{
\begin{fmfgraph}(6,8)
\setval
\fmfleft{v1}
\xvert{v1}
\end{fmfgraph}
}} & $\equiv$ & $\displaystyle{\int\limits_{t_a}^{t_b}dt\,\mean{F(\tx(t))}}$, & 
\parbox{10mm}{\centerline{
\begin{fmfgraph}(6,8)
\setval
\fmfleft{v1}
\pvert{v1}
\end{fmfgraph}
}} & $\equiv$ & $\displaystyle{\int\limits_{t_a}^{t_b}dt\,\mean{F(\tp(t))}}$,\\
\parbox{10mm}{\centerline{
\begin{fmfgraph}(6,8)
\setval
\fmfleft{v1}
\fmfright{v2,v3}
\fmf{plain}{v1,v3}
\xvert{v1}
\end{fmfgraph}
}} & $\equiv$ & $\displaystyle{\int\limits_{t_a}^{t_b}dt\,\mean{F'(\tx(t))}}$, &
\parbox{10mm}{\centerline{
\begin{fmfgraph}(6,8)
\setval
\fmfleft{v1}
\fmfright{v2,v3}
\fmf{wiggly}{v1,v3}
\pvert{v1}
\end{fmfgraph}
}} & $\equiv$ & $\displaystyle{\int\limits_{t_a}^{t_b}dt\,\mean{F'(\tp(t))}}$, \\
\parbox{10mm}{\centerline{
\begin{fmfgraph}(6,8)
\setval
\fmfleft{v1}
\fmfright{v2,v3}
\fmf{plain}{v1,v3}
\fmf{plain}{v1,v2}
\xvert{v1}
\end{fmfgraph}
}} & $\equiv$ & $\displaystyle{\int\limits_{t_a}^{t_b}dt\,\mean{F''(\tx(t))}}$, &
\parbox{10mm}{\centerline{
\begin{fmfgraph}(6,8)
\setval
\fmfleft{v1}
\fmfright{v2,v3}
\fmf{wiggly}{v1,v3}
\fmf{wiggly}{v1,v2}
\pvert{v1}
\end{fmfgraph}
}} & $\equiv$ & $\displaystyle{\int\limits_{t_a}^{t_b}dt\,\mean{F''(\tp(t))}}$, \\
&\vdots &&&\vdots&.
\end{tabular}
\end{center}
With these elements, we can compose Feynman graphs for two-point correlation functions of the type (\ref{der01a}) for arbitrary $n$ by successively applying the generalized Wick rule (\ref{w06c}) or directly using the derivative relations (\ref{der06a})-(\ref{der06d}). The general results become obvious by giving explicitly a graphical representation of the following four
 correlation functions
\begin{eqnarray}
\int\limits_{t_a}^{t_b}dt_1\int\limits_{t_a}^{t_b}dt_2\,\mean{F(\tx(t_1))\,\tx(t_2)}&=&\int\limits_{t_a}^{t_b}dt_1\int\limits_{t_a}^{t_b}dt_2\,iG_{jj}(t_1,t_2)\,\mean{F'(\tx(t_1))}\\
&\equiv&\quad
\parbox{20mm}{\centerline{
\begin{fmfgraph}(15,10)
\setval
\fmfleft{v1}
\fmfright{v2}
\fmf{plain}{v1,v2}
\xvert{v1}
\fmfdot{v2}
\end{fmfgraph}
}},\nonumber\\
\int\limits_{t_a}^{t_b}dt_1\int\limits_{t_a}^{t_b}dt_2\,\mean{F(\tx(t_1))\,\tx^2(t_2)}&=&\int\limits_{t_a}^{t_b}dt_1\int\limits_{t_a}^{t_b}dt_2\,\Bigg\{iG_{jj}(t_2,t_2)\,\mean{F(\tx(t_1))}\nonumber\\
&&+\left[iG_{jj}(t_1,t_2)\right]^2\, \mean{F''(\tx(t_1))}\Bigg\}\\
&\equiv&\quad
\parbox{6mm}{\centerline{
\begin{fmfgraph}(5,5)
\setval
\fmfleft{v1}
\xvert{v1}
\end{fmfgraph}
}}
\parbox{12mm}{\centerline{
\begin{fmfgraph}(8,8)
\setval
\fmfleft{v1}
\fmfright{v2}
\fmf{plain,left}{v1,v2,v1}
\fmfdot{v1}
\end{fmfgraph}
}}
\quad+\quad
\parbox{20mm}{\centerline{
\begin{fmfgraph}(15,15)
\setval
\fmfleft{v1}
\fmfright{v2}
\fmf{plain,left=0.7}{v1,v2,v1}
\xvert{v1}
\fmfdot{v2}
\end{fmfgraph}
}},\nonumber\\
\int\limits_{t_a}^{t_b}dt_1\int\limits_{t_a}^{t_b}dt_2\,\mean{F(\tx(t_1))\,\tx^3(t_2)}&=&\int\limits_{t_a}^{t_b}dt_1\int\limits_{t_a}^{t_b}dt_2\Bigg\{3\,iG_{jj}(t_1,t_2)\,iG_{jj}(t_2,t_2)\mean{F'(\tx(t_1))}\nonumber\\
&&+\left[iG_{jj}(t_1,t_2)\right]^3\, \mean{F'''(\tx(t_1))}\Bigg\}\\
&\equiv&\,3\quad
\parbox{20mm}{\centerline{
\begin{fmfgraph}(16,8)
\setval
\fmfleft{v1}
\fmfright{v3}
\fmf{plain,straight}{v1,v2}
\fmf{plain,left}{v2,v3}
\fmffreeze
\fmf{plain,right}{v2,v3}
\xvert{v1}
\fmfdot{v2}
\end{fmfgraph}
}}\quad+\quad
\parbox{20mm}{\centerline{
\begin{fmfgraph}(16,15)
\setval
\fmfleft{v1}
\fmfright{v2}
\fmf{plain,left=0.7}{v1,v2,v1}
\fmf{plain}{v1,v2}
\xvert{v1}
\fmfdot{v2}
\end{fmfgraph}
}},\nonumber\\
\int\limits_{t_a}^{t_b}dt_1\int\limits_{t_a}^{t_b}dt_2\,\mean{F(\tx(t_1))\,\tx^4(t_2)}&=&\int\limits_{t_a}^{t_b}dt_1\int\limits_{t_a}^{t_b}dt_2\,\Bigg\{\left[iG_{jj}(t_2,t_2)\right]^2\,\mean{F(\tx(t_1))}\nonumber\\
&&+\,6\left[iG_{jj}(t_1,t_2)\right]^2\,iG_{jj}(t_2,t_2)\,\mean{F''(\tx(t_1))}\nonumber\\
&&+\left[iG_{jj}(t_1,t_2)\right]^4\,\mean{F^{(4)}(\tx(t_1))}\Bigg\}\\
&\equiv&\quad
\parbox{6mm}{\centerline{
\begin{fmfgraph}(5,5)
\setval
\fmfleft{v1}
\xvert{v1}
\end{fmfgraph}
}}
\parbox{18mm}{\centerline{
\begin{fmfgraph}(16,8)
\setval
\fmfleft{v1}
\fmfright{v3}
\fmf{plain,left}{v1,v2,v1}
\fmf{plain,left}{v2,v3,v2}
\fmfdot{v2}
\end{fmfgraph}
}}
\;+\,6\;
\parbox{20mm}{\centerline{
\begin{fmfgraph}(16,8)
\setval
\fmfleft{v1}
\fmfright{v3}
\fmf{plain,left}{v1,v2,v1}
\fmf{plain,left}{v2,v3,v2}
\xvert{v1}
\fmfdot{v2}
\end{fmfgraph}
}}\;+\;
\parbox{20mm}{\centerline{
\begin{fmfgraph}(16,15)
\setval
\fmfleft{v1}
\fmfright{v2}
\fmf{plain,left=0.7}{v1,v2,v1}
\fmf{plain,left=0.35}{v1,v2,v1}
\xvert{v1}
\fmfdot{v2}
\end{fmfgraph}
}}.\nonumber
\end{eqnarray}
Mixed position-momentum and momentum-momentum-correlations and their 
graphical representations are given in Appendix~{\ref{corrApp}}.

The consideration of higher-order correlations with more than one 
function $F(\tx(t))$ or $F(\tp(t))$ can be reduced to the results (\ref{w02a})-(\ref{w02d}) or (\ref{der06a})-(\ref{der06d}) by expanding them with respect to the classical path or momentum, respectively. By expanding both functions in the expectation value, one obtains for example
\begin{equation}
  \label{feyn00}
  \mean{F_1(\tx(t_1))\,F_2(\tx(t_2))}=\sum\limits_{m=0}^\infty\,\sum\limits_{n=0}^\infty\,\frac{1}{m!\,n!}\,f_{1,m} f_{2,n}\,\mean{\tx^m(t_1)\,\tx^n(t_2)}
\end{equation}
with
\begin{equation}
  \label{feyn00b}
  f_{i,m}=F^{(m)}(0),\quad i=1,2.
\end{equation}
But, constructing graphical rules for such general correlations is more involved due to the various summations over products of powers of propagators $G_{jj}(t_i,t_j)$ with $i,j=1,2$. 

Finally, we apply the diagrammatic rules to the anharmonic oscillator with 
$\tx^4$-interaction which is a powerful system being discussed in detail by the 
help of perturbation expansion \cite[Chap.~3]{Kleinert}. 
With the Green functions 
given by (\ref{37}), (\ref{58}), and (\ref{59}), the two-point-correlation 
for anharmonic system with arbitrary time-dependent frequency can then be 
expressed graphically, yielding the known decomposition for the second-order perturbative contribution
\begin{eqnarray}
  \label{feyn01}
  \int\limits_{t_a}^{t_b}dt_1\int\limits_{t_a}^{t_b}dt_2\,\cum{\tx^4(t_1)\,\tx^4(t_2)}\equiv\,72\quad
\parbox{25mm}{\centerline{
\begin{fmfgraph}(24,8)
\setval
\fmfleft{v1}
\fmfright{v4}
\fmfforce{0.33w,0.5h}{v2}
\fmfforce{0.66w,0.5h}{v3}
\fmf{plain,left}{v1,v2,v1}
\fmf{plain,left}{v2,v3,v2}
\fmf{plain,left}{v3,v4,v3}
\fmfdot{v2,v3}
\end{fmfgraph}
}}\quad+\,24\quad
\parbox{18mm}{\centerline{
\begin{fmfgraph}(16,15)
\setval
\fmfleft{v1}
\fmfright{v2}
\fmf{plain,left=0.7}{v1,v2,v1}
\fmf{plain,left=0.35}{v1,v2,v1}
\fmfdot{v1,v2}
\end{fmfgraph}
}}
\end{eqnarray}
with subscript $c$ indicating that we restrict to connected graphs only. Beyond this, our theory allows to describe nonstandard systems with polynomial interactions (\ref{f00}) depending on both, position and momentum, to higher-order. Finally, we want to give the graphs for a four-interation $\tx^2\,\tp^2$ to second-order to see the variations of possible graphs in comparison with (\ref{feyn01}):
\begin{eqnarray}
  \label{feyn02}
  &&\int\limits_{t_a}^{t_b}dt_1\int\limits_{t_a}^{t_b}dt_2\,\cum{\tx^2(t_1)\,\tp^2(t_1)\,\tx^2(t_2)\,\tp^2(t_2)}
\equiv\,2\quad
\parbox{25mm}{\centerline{
\begin{fmfgraph}(24,8)
\setval
\fmfleft{v1}
\fmfright{v4}
\fmfforce{0.33w,0.5h}{v2}
\fmfforce{0.66w,0.5h}{v3}
\fmf{wiggly,left}{v1,v2,v1}
\fmf{plain,left}{v2,v3,v2}
\fmf{wiggly,left}{v3,v4,v3}
\fmfdot{v2,v3}
\end{fmfgraph}
}}\quad+\,16\quad
\parbox{25mm}{\centerline{
\begin{fmfgraph}(24,8)
\setval
\fmfleft{v1}
\fmfright{v4}
\fmfforce{0.33w,0.5h}{v2}
\fmfforce{0.66w,0.5h}{v3}
\fmf{wiggly,left}{v1,v2,v1}
\fmf{plain,right}{v2,v3}
\fmf{dashes_arrow,right}{v3,v2}
\fmf{plain,left}{v3,v4,v3}
\fmfdot{v2,v3}
\end{fmfgraph}
}}\nonumber\\
&&\hspace{5mm}+\,16\quad
\parbox{25mm}{\centerline{
\begin{fmfgraph}(24,8)
\setval
\fmfforce{0.33w,0.5h}{v2}
\fmfforce{0.66w,0.5h}{v3}
\fmf{plain,left}{v2,v3}
\fmf{wiggly,left}{v3,v2}
\fmfi{dashes_arrow}{reverse fullcircle scaled 0.33w shifted (0.165w,0.5h)}
\fmfi{dashes_arrow}{fullcircle rotated 180 scaled 0.33w shifted (0.825w,0.5h)}
\fmfdot{v2,v3}
\end{fmfgraph}
}}\:+\,2\quad
\parbox{25mm}{\centerline{
\begin{fmfgraph}(24,8)
\setval
\fmfforce{0.33w,0.5h}{v2}
\fmfforce{0.66w,0.5h}{v3}
\fmf{wiggly,left}{v2,v3}
\fmf{wiggly,left}{v3,v2}
\fmfi{plain}{reverse fullcircle scaled 0.33w shifted (0.165w,0.5h)}
\fmfi{plain}{fullcircle rotated 180 scaled 0.33w shifted (0.825w,0.5h)}
\fmfdot{v2,v3}
\end{fmfgraph}
}}\:+\,4\quad
\parbox{25mm}{\centerline{
\begin{fmfgraph}(24,8)
\setval
\fmfforce{0.33w,0.5h}{v2}
\fmfforce{0.66w,0.5h}{v3}
\fmf{dashes_arrow,left}{v2,v3}
\fmf{dashes_arrow,right}{v2,v3}
\fmfi{plain}{reverse fullcircle scaled 0.33w shifted (0.165w,0.5h)}
\fmfi{wiggly}{fullcircle rotated 180 scaled 0.33w shifted (0.825w,0.5h)}
\fmfdot{v2,v3}
\end{fmfgraph}
}}\:+\,16\quad
\parbox{25mm}{\centerline{
\begin{fmfgraph}(24,8)
\setval
\fmfforce{0.33w,0.5h}{v2}
\fmfforce{0.66w,0.5h}{v3}
\fmf{dashes_arrow,left}{v2,v3}
\fmf{wiggly,right}{v2,v3}
\fmfi{plain}{fullcircle scaled 0.33w shifted (0.165w,0.5h)}
\fmfi{dashes_arrow}{fullcircle rotated 180 scaled 0.33w shifted (0.825w,0.5h)}
\fmfdot{v2,v3}
\end{fmfgraph}
}}\nonumber\\
&&\hspace{5mm}+\,16\quad
\parbox{25mm}{\centerline{
\begin{fmfgraph}(24,8)
\setval
\fmfforce{0.33w,0.5h}{v2}
\fmfforce{0.66w,0.5h}{v3}
\fmf{dashes_arrow,left}{v2,v3}
\fmf{dashes_arrow,left}{v3,v2}
\fmfi{dashes_arrow}{reverse fullcircle scaled 0.33w shifted (0.165w,0.5h)}
\fmfi{dashes_arrow}{fullcircle rotated 180 scaled 0.33w shifted (0.825w,0.5h)}
\fmfdot{v2,v3}
\end{fmfgraph}
}}\quad+\,4\quad
\parbox{18mm}{\centerline{
\begin{fmfgraph}(16,15)
\setval
\fmfleft{v1}
\fmfright{v2}
\fmf{dashes_arrow,left=0.7}{v1,v2,v1}
\fmf{dashes_arrow,left=0.35}{v1,v2,v1}
\fmfdot{v1,v2}
\end{fmfgraph}
}}\quad+\,16\quad
\parbox{18mm}{\centerline{
\begin{fmfgraph}(16,15)
\setval
\fmfleft{v1}
\fmfright{v2}
\fmf{wiggly,left=0.7}{v1,v2}
\fmf{plain,right=0.7}{v1,v2}
\fmf{dashes_arrow,left=0.35}{v1,v2,v1}
\fmfdot{v1,v2}
\end{fmfgraph}
}}\quad+\,4\quad
\parbox{18mm}{\centerline{
\begin{fmfgraph}(16,15)
\setval
\fmfleft{v1}
\fmfright{v2}
\fmf{wiggly,left=0.7}{v1,v2,v1}
\fmf{plain,left=0.35}{v1,v2,v1}
\fmfdot{v1,v2}
\end{fmfgraph}
}}\;.
\end{eqnarray}
We see, that we have the same class of graphs already occuring in (\ref{feyn01}), however, with different propagators connecting the vertices. Thus, both classes decay into subclasses with different multiplicities, but the total numbers remain 72 and 24 for each type of class, respectively. Furthermore, all graphs are vacuum-like graphs. Eventually, it is easy to construct the Feynman graphs for polynomial correlations higher than second order by applying Wick's rule or the Feynman rules given in this section.
\section{Simplifications for Periodic Paths}
\label{genper}
Up to now, we discussed the harmonic time evolution amplitude with arbitrary frequency and external sources $j(t),k(t)$ and corresponding Green functions fulfilling Dirichlet boundary conditions. In the sense of the quantum mechanical partition function
\begin{equation}
  \label{per01}
  Z=\int\limits_{-\infty}^{+\infty}dx\,(x\,t_b\,|\,x\,t_a),
\end{equation}
which is an integral over the time evolution amplitude for closed paths, it is of interest to investigate the generating functional for closed paths. In analogy to (\ref{per01}), we define 
\begin{equation}
  \label{per02}
  Z[j,k]=\int\limits_{-\infty}^{+\infty}dx\,(x\,t_b\,|\,x\,t_a)[j,k]
\end{equation}
with (\ref{57}) for $x_a=x_b=x$. One immediatly observes that $Z=Z[0,0]$. The integral is easily done, giving 
\begin{eqnarray}
  \label{per03}
  Z[j,k]&=&\frac{1}{\sqrt{\dot{D}_a(t_b)-\dot{D}_b(t_a)-2}}\,\exp\Bigg\{-\frac{i}{2\hbar}\int\limits_{t_a}^{t_b}dt\int\limits_{t_a}^{t_b}dt'\Bigg[\frac{1}{M}j(t)\,\tilde{G}_{jj}^{x}(t,t')\,j(t') \nonumber\\
&&+\, j(t)\,\tilde{G}_{jk}^{x}(t,t')\,k(t') + k(t)\,\tilde{G}_{kj}^{x}(t,t')\,j(t') + Mk(t)\,\tilde{G}_{kk}^{x}(t,t')\,k(t') \Bigg]\Bigg\}.
\end{eqnarray}
The Green functions, expressed with fundamental solutions (\ref{19}), (\ref{20}), are found to be
\begin{eqnarray}
  \label{per04}
  \tilde{G}_{jj}^{x}(t,t')&=&\frac{1}{D_a(t_b)}\left[G_{jj}^x(t,t')+\frac{1}{a(t_a,t_b)}\,g(t)\,g(t') \right],\\
  \tilde{G}_{jk}^{x}(t,t')&=&\frac{1}{D_a(t_b)}\left[G_{jk}^x(t,t')+\frac{1}{a(t_a,t_b)}\,g(t)\,\dot{g}(t') \right]=\tilde{G}_{kj}^{x}(t',t),\\
  \tilde{G}_{kk}^{x}(t,t')&=&\frac{1}{D_a(t_b)}\left[G_{kk}^x(t,t')+\frac{1}{a(t_a,t_b)}\,\dot{g}(t)\,\dot{g}(t') \right],
\end{eqnarray}
with
\begin{equation}
  \label{per05}
  a(t,t')=\dot{D}_a(t')-\dot{D}_b(t)-2.
\end{equation}
Since the function 
\begin{equation}
  \label{per06}
  g(t)=D_a(t)+D_b(t)
\end{equation}
is periodic, $g(t_a)=g(t_b)$, due to conditions (\ref{19}), (\ref{20}), and (\ref{22}), also the Green function $\tilde{G}_{jj}^{x}(t,t')$ becomes periodic,
\begin{equation}
  \label{per07}
  \tilde{G}_{jj}^{x}(t_a,t')=\tilde{G}_{jj}^{x}(t_b,t').
\end{equation}
In analogy to the harmonic propagator without external sources (\ref{sm01}) we can define expectation values consisting of $N$ position-dependent functions and $M$ momentum-dependent functions by
\begin{eqnarray}
  \label{per08}
  &&\langle\,F_1(x(t_1))\,F_2(x(t_2))\cdots F_{N+M}(p(t_{N+M}))\,\rangle_\Omega=\nonumber\\
&&\hspace{10mm}\frac{1}{Z}\oint\frac{{\cal D}x\,{\cal D}p}{2\pi\hbar}\,F_1(x(t_1))\,F_2(x(t_2))\cdots F_{N+M}(p(t_{N+M}))\,\exp\left\{ \frac{i}{\hbar}{\cal A}[p,x;0,0]\right\}.
\end{eqnarray}
We remark that the generalization of Wick's rule and the graphical representation with the help of Feynman diagrams of such correlation functions is exactly the same as given in the last section after substituting the Green functions $G(t,t')$ by $\tilde{G}(t,t')$ and expectation values (\ref{sm01}) by (\ref{per08}). 
\section{Summary and Outlook}
\label{summary}
We have reduced generating functionals with fixed end points to those with vanishing end points by adding special singular sources to the currents. The new generating functionals were calculated explicitly for the harmonic oscillator with time-dependent frequency. From this expression, a smearing formula was derived which serves to calculate correlation functions for arbitrary polynomial or nonpolynomial position- and momentum-dependent couplings. We have further found a generalization of Wick's theorem of decomposing correlation functions involving functions of the canonic variables of the system. This gives rise to certain generalized Feynman rules for position- and momentum-dependent expectation values. 

Due to its universality, the theory should serve as a basis for investigating physical systems with nonstandard Hamiltonian via perturbation theory and its variational extension. Note, that a perturbation theory for momentum-dependent interactions arises in important field theories such as the nonlinear $\sigma$-model. Our work is supposed to prepare the grounds for a more efficient perturbation treatment of such theory. 
\section*{Acknowledgements}
One of us (M.B.) is supported by the Studienstiftung des deutschen Volkes. 
\begin{appendix}
\section{Generalized Correlation Functions}
\label{corrApp}
In this appendix we give the expectations for the correlation between a general position or momentum dependent function and a polynomial up to order $n=4$:\par\noindent
{\bf Position-Momentum-Coupling:}
\begin{eqnarray}
\int\limits_{t_a}^{t_b}dt_1\int\limits_{t_a}^{t_b}dt_2\,\mean{F(\tx(t_1))\,\tp(t_2)}&=&\int\limits_{t_a}^{t_b}dt_1\int\limits_{t_a}^{t_b}dt_2\,iG_{jk}(t_1,t_2)\,\mean{F'(\tx(t_1))}\\
&\equiv&\quad
\parbox{20mm}{\centerline{
\begin{fmfgraph}(15,10)
\setval
\fmfleft{v1}
\fmfright{v2}
\fmf{dashes_arrow}{v2,v1}
\xvert{v1}
\fmfdot{v2}
\end{fmfgraph}
}},\nonumber\\
\int\limits_{t_a}^{t_b}dt_1\int\limits_{t_a}^{t_b}dt_2\,\mean{F(\tx(t_1))\,\tp^2(t_2)}&=&\int\limits_{t_a}^{t_b}dt_1\int\limits_{t_a}^{t_b}dt_2\,\Bigg\{iG_{kk}(t_2,t_2)\,\mean{F(\tx(t_1))}\nonumber\\
&&+\left[iG_{jk}(t_1,t_2)\right]^2\, \mean{F''(\tx(t_1))}\Bigg\}\\
&\equiv&\quad
\parbox{6mm}{\centerline{
\begin{fmfgraph}(5,5)
\setval
\fmfleft{v1}
\xvert{v1}
\end{fmfgraph}
}}
\parbox{12mm}{\centerline{
\begin{fmfgraph}(8,8)
\setval
\fmfleft{v1}
\fmfright{v2}
\fmf{wiggly,left}{v1,v2,v1}
\fmfdot{v1}
\end{fmfgraph}
}}
\quad+\quad
\parbox{20mm}{\centerline{
\begin{fmfgraph}(15,15)
\setval
\fmfleft{v1}
\fmfright{v2}
\fmf{dashes_arrow,left=0.7}{v2,v1}
\fmf{dashes_arrow,right=0.7}{v2,v1}
\xvert{v1}
\fmfdot{v2}
\end{fmfgraph}
}},\nonumber\\
\int\limits_{t_a}^{t_b}dt_1\int\limits_{t_a}^{t_b}dt_2\,\mean{F(\tx(t_1))\,\tp^3(t_2)}&=&\int\limits_{t_a}^{t_b}dt_1\int\limits_{t_a}^{t_b}dt_2\Bigg\{3\,iG_{jk}(t_1,t_2)\, iG_{kk}(t_2,t_2)\mean{F'(\tx(t_1))}\nonumber\\
&&+\left[iG_{jk}(t_1,t_2)\right]^3\, \mean{F'''(\tx(t_1))}\Bigg\}\\
&\equiv&\,3\quad
\parbox{24mm}{\centerline{
\begin{fmfgraph}(20,8)
\setval
\fmfleft{v1}
\fmfright{v3}
\fmfforce{0.6w,0.5h}{v2}
\fmf{dashes_arrow}{v2,v1}
\fmf{wiggly,left}{v2,v3}
\fmf{wiggly,right}{v2,v3}
\xvert{v1}
\fmfdot{v2}
\end{fmfgraph}
}}\quad+\quad
\parbox{20mm}{\centerline{
\begin{fmfgraph}(16,15)
\setval
\fmfleft{v1}
\fmfright{v2}
\fmf{dashes_arrow,left=0.7}{v2,v1}
\fmf{dashes_arrow,straight}{v2,v1}
\fmf{dashes_arrow,right=0.7}{v2,v1}
\xvert{v1}
\fmfdot{v2}
\end{fmfgraph}
}},\nonumber\\
\int\limits_{t_a}^{t_b}dt_1\int\limits_{t_a}^{t_b}dt_2\,\mean{F(\tx(t_1))\,\tp^4(t_2)}&=&\int\limits_{t_a}^{t_b}dt_1\int\limits_{t_a}^{t_b}dt_2\,\Bigg\{\left[iG_{kk}(t_2,t_2)\right]^2\,\mean{F(\tx(t_1))}\nonumber\\
&&+\,6\left[iG_{jk}(t_1,t_2)\right]^2\,iG_{kk}(t_2,t_2)\,\mean{F''(\tx(t_1))}\nonumber\\
&&+\left[iG_{jk}(t_1,t_2)\right]^4\,\mean{F^{(4)}(\tx(t_1))}\Bigg\}\\
&\equiv&\quad
\parbox{6mm}{\centerline{
\begin{fmfgraph}(5,5)
\setval
\fmfleft{v1}
\xvert{v1}
\end{fmfgraph}
}}
\parbox{18mm}{\centerline{
\begin{fmfgraph}(16,8)
\setval
\fmfleft{v1}
\fmfright{v3}
\fmf{wiggly,left}{v1,v2,v1}
\fmf{wiggly,left}{v3,v2,v3}
\fmfdot{v2}
\end{fmfgraph}
}}
\;+\,6\;
\parbox{24mm}{\centerline{
\begin{fmfgraph}(20,8)
\setval
\fmfleft{v1}
\fmfright{v3}
\fmfforce{0.6w,0.5h}{v2}
\fmf{dashes_arrow,left=0.7}{v2,v1}
\fmf{dashes_arrow,right=0.7}{v2,v1}
\fmf{wiggly,left}{v2,v3,v2}
\xvert{v1}
\fmfdot{v2}
\end{fmfgraph}
}}\;+\;
\parbox{20mm}{\centerline{
\begin{fmfgraph}(16,15)
\setval
\fmfleft{v1}
\fmfright{v2}
\fmf{dashes_arrow,left=0.7}{v2,v1}
\fmf{dashes_arrow,left=0.35}{v2,v1}
\fmf{dashes_arrow,right=0.7}{v2,v1}
\fmf{dashes_arrow,right=0.35}{v2,v1}
\xvert{v1}
\fmfdot{v2}
\end{fmfgraph}
}};\nonumber
\end{eqnarray}
{\bf Momentum-Position-Coupling:}
\begin{eqnarray}
\int\limits_{t_a}^{t_b}dt_1\int\limits_{t_a}^{t_b}dt_2\,\mean{F(\tp(t_1))\,\tx(t_2)}&=&\int\limits_{t_a}^{t_b}dt_1\int\limits_{t_a}^{t_b}dt_2\,iG_{kj}(t_1,t_2)\,\mean{F'(\tp(t_1))}\\
&\equiv&\quad
\parbox{20mm}{\centerline{
\begin{fmfgraph}(15,10)
\setval
\fmfleft{v1}
\fmfright{v2}
\fmf{dashes_arrow}{v1,v2}
\pvert{v1}
\fmfdot{v2}
\end{fmfgraph}
}},\nonumber\\
\int\limits_{t_a}^{t_b}dt_1\int\limits_{t_a}^{t_b}dt_2\,\mean{F(\tp(t_1))\,\tx^2(t_2)}&=&\int\limits_{t_a}^{t_b}dt_1\int\limits_{t_a}^{t_b}dt_2\,\Bigg\{iG_{jj}(t_2,t_2)\,\mean{F(\tp(t_1))}\nonumber\\
&&+\left[iG_{kj}(t_1,t_2)\right]^2\, \mean{F''(\tp(t_1))}\Bigg\}\\
&\equiv&\quad
\parbox{6mm}{\centerline{
\begin{fmfgraph}(5,5)
\setval
\fmfleft{v1}
\pvert{v1}
\end{fmfgraph}
}}
\parbox{12mm}{\centerline{
\begin{fmfgraph}(8,8)
\setval
\fmfleft{v1}
\fmfright{v2}
\fmf{plain,left}{v1,v2,v1}
\fmfdot{v1}
\end{fmfgraph}
}}
\quad+\quad
\parbox{20mm}{\centerline{
\begin{fmfgraph}(15,15)
\setval
\fmfleft{v1}
\fmfright{v2}
\fmf{dashes_arrow,left=0.7}{v1,v2}
\fmf{dashes_arrow,right=0.7}{v1,v2}
\pvert{v1}
\fmfdot{v2}
\end{fmfgraph}
}},\nonumber\\
\int\limits_{t_a}^{t_b}dt_1\int\limits_{t_a}^{t_b}dt_2\,\mean{F(\tp(t_1))\,\tx^3(t_2)}&=&\int\limits_{t_a}^{t_b}dt_1\int\limits_{t_a}^{t_b}dt_2\Bigg\{3\,iG_{kj}(t_1,t_2)\, iG_{jj}(t_2,t_2)\mean{F'(\tp(t_1))}\nonumber\\
&&+\left[iG_{kj}(t_1,t_2)\right]^3\, \mean{F'''(\tp(t_1))}\Bigg\}\\
&\equiv&\,3\quad
\parbox{24mm}{\centerline{
\begin{fmfgraph}(20,8)
\setval
\fmfleft{v1}
\fmfright{v3}
\fmfforce{0.6w,0.5h}{v2}
\fmf{dashes_arrow}{v1,v2}
\fmf{plain,left}{v2,v3}
\fmf{plain,right}{v2,v3}
\pvert{v1}
\fmfdot{v2}
\end{fmfgraph}
}}\quad+\quad
\parbox{20mm}{\centerline{
\begin{fmfgraph}(16,15)
\setval
\fmfleft{v1}
\fmfright{v2}
\fmf{dashes_arrow,left=0.7}{v1,v2}
\fmf{dashes_arrow,straight}{v1,v2}
\fmf{dashes_arrow,right=0.7}{v1,v2}
\pvert{v1}
\fmfdot{v2}
\end{fmfgraph}
}},\nonumber\\
\int\limits_{t_a}^{t_b}dt_1\int\limits_{t_a}^{t_b}dt_2\,\mean{F(\tp(t_1))\,\tx^4(t_2)}&=&\int\limits_{t_a}^{t_b}dt_1\int\limits_{t_a}^{t_b}dt_2\,\Bigg\{\left[iG_{jj}(t_2,t_2)\right]^2\,\mean{F(\tp(t_1))}\nonumber\\
&&+\,6\left[iG_{kj}(t_1,t_2)\right]^2\,iG_{jj}(t_2,t_2)\,\mean{F''(\tp(t_1))}\nonumber\\
&&+\left[iG_{kj}(t_1,t_2)\right]^4\,\mean{F^{(4)}(\tp(t_1))}\Bigg\}\\
&\equiv&\quad
\parbox{6mm}{\centerline{
\begin{fmfgraph}(5,5)
\setval
\fmfleft{v1}
\pvert{v1}
\end{fmfgraph}
}}
\parbox{18mm}{\centerline{
\begin{fmfgraph}(16,8)
\setval
\fmfleft{v1}
\fmfright{v3}
\fmf{plain,left}{v1,v2,v1}
\fmf{plain,left}{v3,v2,v3}
\fmfdot{v2}
\end{fmfgraph}
}}
\;+\,6\;
\parbox{24mm}{\centerline{
\begin{fmfgraph}(20,8)
\setval
\fmfleft{v1}
\fmfright{v3}
\fmfforce{0.6w,0.5h}{v2}
\fmf{dashes_arrow,left=0.7}{v1,v2}
\fmf{dashes_arrow,right=0.7}{v1,v2}
\fmf{plain,left}{v2,v3,v2}
\pvert{v1}
\fmfdot{v2}
\end{fmfgraph}
}}\;+\;
\parbox{20mm}{\centerline{
\begin{fmfgraph}(16,15)
\setval
\fmfleft{v1}
\fmfright{v2}
\fmf{dashes_arrow,left=0.7}{v1,v2}
\fmf{dashes_arrow,left=0.35}{v1,v2}
\fmf{dashes_arrow,right=0.7}{v1,v2}
\fmf{dashes_arrow,right=0.35}{v1,v2}
\pvert{v1}
\fmfdot{v2}
\end{fmfgraph}
}};\nonumber
\end{eqnarray}
{\bf Momentum-Momentum-Coupling:}
\begin{eqnarray}
\int\limits_{t_a}^{t_b}dt_1\int\limits_{t_a}^{t_b}dt_2\,\mean{F(\tp(t_1))\,\tp(t_2)}&=&\int\limits_{t_a}^{t_b}dt_1\int\limits_{t_a}^{t_b}dt_2\,iG_{kk}(t_1,t_2)\,\mean{F'(\tp(t_1))}\\
&\equiv&\quad
\parbox{20mm}{\centerline{
\begin{fmfgraph}(15,10)
\setval
\fmfleft{v1}
\fmfright{v2}
\fmf{wiggly}{v1,v2}
\pvert{v1}
\fmfdot{v2}
\end{fmfgraph}
}},\nonumber\\
\int\limits_{t_a}^{t_b}dt_1\int\limits_{t_a}^{t_b}dt_2\,\mean{F(\tp(t_1))\,\tp^2(t_2)}&=&\int\limits_{t_a}^{t_b}dt_1\int\limits_{t_a}^{t_b}dt_2\,\Bigg\{iG_{kk}(t_2,t_2)\,\mean{F(\tp(t_1))}\nonumber\\
&&+\left[iG_{kk}(t_1,t_2)\right]^2\, \mean{F''(\tp(t_1))}\Bigg\}\\
&\equiv&\quad
\parbox{6mm}{\centerline{
\begin{fmfgraph}(5,5)
\setval
\fmfleft{v1}
\pvert{v1}
\end{fmfgraph}
}}
\parbox{12mm}{\centerline{
\begin{fmfgraph}(8,8)
\setval
\fmfleft{v1}
\fmfright{v2}
\fmf{wiggly,left}{v1,v2,v1}
\fmfdot{v1}
\end{fmfgraph}
}}
\quad+\quad
\parbox{20mm}{\centerline{
\begin{fmfgraph}(15,15)
\setval
\fmfleft{v1}
\fmfright{v2}
\fmf{wiggly,left=0.7}{v1,v2}
\fmf{wiggly,right=0.7}{v1,v2}
\pvert{v1}
\fmfdot{v2}
\end{fmfgraph}
}},\nonumber\\
\int\limits_{t_a}^{t_b}dt_1\int\limits_{t_a}^{t_b}dt_2\,\mean{F(\tp(t_1))\,\tp^3(t_2)}&=&\int\limits_{t_a}^{t_b}dt_1\int\limits_{t_a}^{t_b}dt_2\Bigg\{3\,iG_{kk}(t_1,t_2)\, iG_{kk}(t_2,t_2)\mean{F'(\tp(t_1))}\nonumber\\
&&+\left[iG_{kk}(t_1,t_2)\right]^3\, \mean{F'''(\tp(t_1))}\Bigg\}\\
&\equiv&\,3\quad
\parbox{20mm}{\centerline{
\begin{fmfgraph}(16,8)
\setval
\fmfleft{v1}
\fmfright{v3}
\fmf{wiggly}{v1,v2}
\fmf{wiggly,left}{v2,v3}
\fmffreeze
\fmf{wiggly,right}{v2,v3}
\pvert{v1}
\fmfdot{v2}
\end{fmfgraph}
}}\quad+\quad
\parbox{20mm}{\centerline{
\begin{fmfgraph}(16,15)
\setval
\fmfleft{v1}
\fmfright{v2}
\fmf{wiggly,left=0.7}{v1,v2}
\fmf{wiggly,straight}{v1,v2}
\fmf{wiggly,right=0.7}{v1,v2}
\pvert{v1}
\fmfdot{v2}
\end{fmfgraph}
}},\nonumber\\
\int\limits_{t_a}^{t_b}dt_1\int\limits_{t_a}^{t_b}dt_2\,\mean{F(\tp(t_1))\,\tp^4(t_2)}&=&\int\limits_{t_a}^{t_b}dt_1\int\limits_{t_a}^{t_b}dt_2\,\Bigg\{\left[iG_{kk}(t_2,t_2)\right]^2\,\mean{F(\tp(t_1))}\nonumber\\
&&+\,6\left[iG_{kk}(t_1,t_2)\right]^2\,iG_{kk}(t_2,t_2)\,\mean{F''(\tp(t_1))}\nonumber\\
&&+\left[iG_{kk}(t_1,t_2)\right]^4\,\mean{F^{(4)}(\tp(t_1))}\Bigg\}\\
&\equiv&\quad
\parbox{6mm}{\centerline{
\begin{fmfgraph}(5,5)
\setval
\fmfleft{v1}
\pvert{v1}
\end{fmfgraph}
}}
\parbox{18mm}{\centerline{
\begin{fmfgraph}(16,8)
\setval
\fmfleft{v1}
\fmfright{v3}
\fmf{wiggly,left}{v1,v2,v1}
\fmf{wiggly,left}{v3,v2,v3}
\fmfdot{v2}
\end{fmfgraph}
}}
\;+\,6\;
\parbox{20mm}{\centerline{
\begin{fmfgraph}(16,8)
\setval
\fmfleft{v1}
\fmfright{v3}
\fmf{wiggly,left}{v1,v2,v1}
\fmf{wiggly,left}{v2,v3,v2}
\pvert{v1}
\fmfdot{v2}
\end{fmfgraph}
}}\;+\;
\parbox{20mm}{\centerline{
\begin{fmfgraph}(16,15)
\setval
\fmfleft{v1}
\fmfright{v2}
\fmf{wiggly,left=0.7}{v1,v2}
\fmf{wiggly,left=0.35}{v1,v2}
\fmf{wiggly,right=0.7}{v1,v2}
\fmf{wiggly,right=0.35}{v1,v2}
\pvert{v1}
\fmfdot{v2}
\end{fmfgraph}
}}.\nonumber
\end{eqnarray}
The case of position-position-coupling has already been calculated in Sect.~{\ref{feynman}}.

\end{appendix}
\end{fmffile}
\end{document}